\documentclass[aip,cha,reprint,amsmath,amssymb,nofootinbib]{revtex4-2}

\usepackage{graphicx}
\usepackage{dcolumn}
\usepackage{bm}
\usepackage{dsfont}
\usepackage{booktabs}
\usepackage{xcolor}
\usepackage{subcaption}
\usepackage{tikz}
\usepackage{pgfplots}
\pgfplotsset{compat=1.16}
\usepackage[colorlinks=true,allcolors=blue]{hyperref}

\usepackage{mathtools}
\mathtoolsset{showonlyrefs=true}

\newcommand{\EE}{\mathbb{E}}
\newcommand{\RR}{\mathbb{R}}
\newcommand{\PP}{\mathbb{P}}
\newcommand{\eps}{\varepsilon}
\newcommand{\dd}{\mathrm{d}}
\DeclareMathOperator{\Id}{Id}
\DeclareMathOperator{\ppr}{pr}
\DeclareMathOperator{\dom}{dom}
\DeclareMathOperator{\trace}{tr}

\newcommand{\norm}[1]{\left\lVert#1\right\rVert}

\makeatletter
\renewcommand\@makecaption[2]{%
  \par
  \vskip\abovecaptionskip
  \begingroup
   \small\rmfamily
    \begingroup
     \samepage
     \flushing
     \let\footnote\@footnotemark@gobble
     \@make@capt@title{#1}{#2}\par
    \endgroup
  \endgroup
  \vskip\belowcaptionskip
}
\makeatother

\usepackage{orcidlink}

\begin{document}

\title[Computing rare events in L{\'e}vy-driven systems via optimal control]{Computing rare events in L{\'e}vy-driven systems via optimal control}

\author{Jinan Pan\,\orcidlink{0009-0002-8372-6263}}
\email{andypan@uchicago.edu}
\affiliation{Committee on Computational and Applied Mathematics, University of Chicago, Chicago, IL 60637, USA}

\author{Timo Schorlepp\,\orcidlink{0000-0002-9143-8854}}
\email{timo.schorlepp@nyu.edu}
\affiliation{Courant Institute School of Mathematics, Computing, and Data Science, New York University, New York, NY 10012, USA}

\date{\today}

\begin{abstract}
We extend the optimal control formulation of sample path large deviation theory, which has proven effective for numerical computation of instantons and sharp probability estimates in stochastic differential equations with Gaussian noise, to systems with light-tailed L{\'e}vy noise. In the scaling limit of small jumps and high intensity, such systems satisfy a large deviation principle of Freidlin--Wentzell type, with a running cost given by the Legendre transform of the noise cumulant generating function. We propose to numerically minimize this cost over noise realizations subject to the system dynamics, which yields the instanton without having to solve Hamilton's equations, and naturally handles degenerate and multiplicative noise. When the cumulant generating function has a bounded domain, the cost grows only linearly for large noise, and instantons can contain jumps. We analyze when jumps are not only admissible but optimal, depending on the subleading behavior of the cost and on the drift, and show how jumps can be resolved numerically through a time reparameterization that freezes physical time. For continuous instantons, we derive Riccati and Fredholm determinant expressions for the leading-order prefactor, and we discuss the modifications needed when the instanton jumps. We illustrate and validate our methods against Monte Carlo simulations in several examples: a two-dimensional nonlinear model, an active run-and-tumble particle with non-Gaussian tumbling noise, and the Ghil--Sellers energy balance climate model with L{\'e}vy-perturbed solar forcing.
\end{abstract}

\maketitle

\begin{quotation}
Rare but important events in noisy dynamical systems, such as transitions between climate states or large excursions of an active particle in a trap, are often dominated by a single most likely pathway, the instanton. For systems driven by Gaussian white noise, instantons and accurate estimates of event probabilities can be computed efficiently by viewing the noise as a control and minimizing its cost subject to the system dynamics. Many natural and engineered systems, however, are subject to noise with jumps, such as random kicks, tumbles, or bursts of forcing. Here, we show that the same control viewpoint extends to systems driven by L{\'e}vy noise with jump size distributions with exponential or faster-decaying tails, in the regime of many small jumps. In this regime, extreme events can be exponentially more likely than under Gaussian noise of the same variance. An important feature is that the optimal path itself may contain jumps, which we characterize and resolve numerically by artificially stretching time. We further show how to refine the resulting probability estimate by a computable prefactor, and test our methods and predictions on examples of increasing complexity, including a climate model.
\end{quotation}

%%%%%%%%%%%%%%%%%%%%%%%%%%%%%%%%%%%%%%%%%%%%%%%%%%%%%%%%%%%%%%%%%%%%%%%%%%%
\section{Introduction}
\label{sec:intro}
%%%%%%%%%%%%%%%%%%%%%%%%%%%%%%%%%%%%%%%%%%%%%%%%%%%%%%%%%%%%%%%%%%%%%%%%%%%
Sample path large deviation theory (LDT)~\cite{freidlin-wentzell:2012,dembo-zeitouni:1998} describes the probability of rare events in stochastic systems with small noise through a variational problem. As the noise strength $\eps \downarrow 0$, the probability of an event decays as $\exp\{-I/\eps\}$, where $I$, the rate function of the event, is the minimal action over all paths that realize the event. The minimizer, called the instanton, is the most likely realization of the event, and often dominates its probability. Instantons thus provide both the leading-order probability of a rare event and its most likely mechanism, and their numerical computation has become a standard tool in physics, chemistry, fluid dynamics and climate science~\cite{grafke-vanden-eijnden:2019}. Established algorithms include the minimum action method and its geometric variant (gMAM)~\cite{e-ren-vanden-eijnden:2004,heymann-vanden-eijnden:2008}, as well as iterative solvers of the associated Hamiltonian boundary value problem~\cite{chernykh-stepanov:2001,grafke-grauer-schaefer:2015}.

\begin{figure*}
\centering
\includegraphics[width=\textwidth]{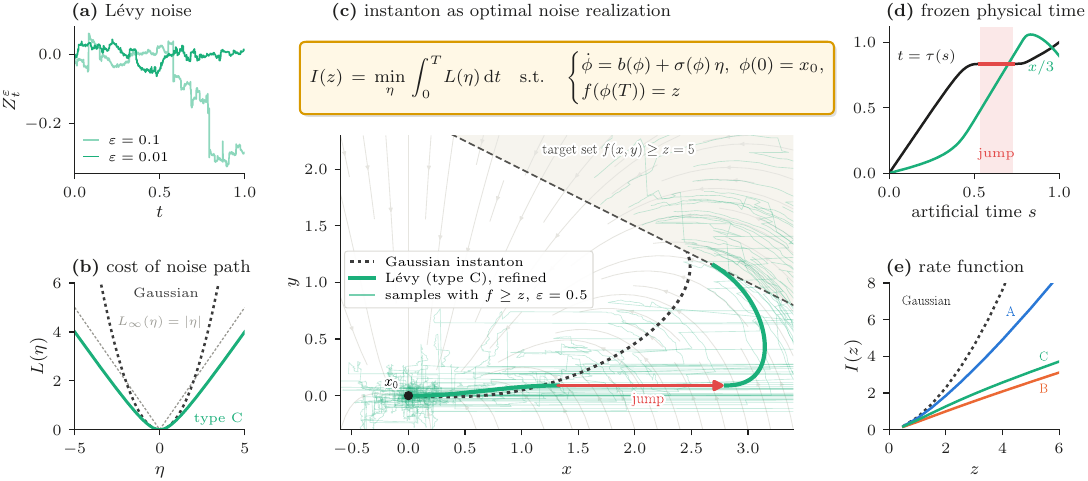}
\caption{Overview, illustrated with the two-dimensional model of Sec.~\ref{sec:2d}. (a)~Sample paths of the rescaled (type C) L{\'e}vy process $Z^\eps_t = \eps Z_{t/\eps}$ (see Sec.~\ref{sec:noise-types}). As $\eps \downarrow 0$, jumps become small and frequent, and typical paths look diffusive. (b)~Cost $L(\eta)$ of a noise realization, which is quadratic for Gaussian noise, but grows only linearly for jump sizes with exponential tails, such that large noise is comparatively cheap. The thin dashed line shows the recession cost $L_\infty(\eta)=|\eta|$. (c)~The instanton for the rare event $\{f(X^\eps_T) \geq z\}$ with $z=5$ is the cheapest noise realization that drives the system into the shaded target set, see the boxed optimal control problem or Eq.~\eqref{eq:control-problem}. The background shows the drift field. For type C noise, the instanton contains a jump, marked by the red arrow, as do Monte Carlo samples conditioned on the event, shown as thin green lines. (d)~The underlying time-reparameterized OC solution freezes physical time $t = \tau(s)$ while the state moves. (e)~Rate functions of $(f(X_T^\eps))_{\eps > 0}$ for Gaussian noise and for noise of types A, B and C ($\alpha = 1/2$), see Table~\ref{tab:noise-types}, all with the same covariance: extreme events are exponentially more likely for L{\'e}vy noise than for Gaussian noise.}
\label{fig:concept}
\end{figure*}

For stochastic differential equations (SDEs) driven by Gaussian noise, it is convenient to formulate the instanton problem as an optimal control (OC) problem~\cite{grafke-vanden-eijnden:2019,schorlepp-tong-grafke-etal:2023,de-giuli:2025,schorlepp-grafke:2025,wu-chao-duan:2026}. The noise realization then acts as the control, and its quadratic cost is minimized subject to the system dynamics. This viewpoint has several practical advantages~\cite{schorlepp-tong-grafke-etal:2023,schorlepp-grafke:2025}: the optimization is unconstrained apart from the target event, gradients are obtained by an adjoint solve (or automatic differentiation), degenerate or multiplicative noise requires no special treatment, and existing PDE-constrained optimization machinery can be used. Furthermore, the OC formulation leads directly to a sharp estimate of the event probability including its pre-exponential factor, expressed through a Fredholm determinant~\cite{schorlepp-tong-grafke-etal:2023,schorlepp-grafke:2025}. This determinant can be evaluated matrix-free via dominant eigenvalues and is hence easily scalable to high dimensions. Equivalent expressions for the prefactor in terms of matrix Riccati equations exist~\cite{maier-stein:1996,schorlepp-grafke-grauer:2021,bouchet-reygner:2022,schorlepp-grafke-grauer:2023,grafke-schaefer-vanden-eijnden:2023}, but require more effort to scale to high dimensions.

Many systems of interest are driven by noise that is not Gaussian and contains jumps, from jump-like extreme events in climate forcing~\cite{ditlevsen:1999,sura-newman-penland-etal:2005,lucarini-serdukova-margazoglou:2022} to tumbling of active particles~\cite{tailleur-cates:2008} and anomalous transport~\cite{metzler-klafter:2000}. Such noise is naturally modeled by more general L{\'e}vy processes~\cite{sato:1999,applebaum:2009} instead of Brownian motion. For heavy-tailed, for instance $\alpha$-stable, L{\'e}vy noise, rare events are typically caused by a single large jump, and transition rates scale polynomially rather than exponentially in the noise strength~\cite{imkeller-pavlyukevich:2006,imkeller-pavlyukevich-wetzel:2010}. In this paper, we instead consider \emph{light-tailed} L{\'e}vy noise, for which the cumulant generating function (CGF) of the noise is finite at least on a neighborhood of the origin, in the scaling limit in which jumps become small while their intensity increases. In this limit, which is natural when many weak, independent kicks act on a system, the noise behaves diffusively at the level of typical fluctuations, but its large deviations retain the full information on the jump distribution. The corresponding sample path large deviation principle (LDP) is still of Freidlin--Wentzell type~\cite{freidlin-wentzell:2012,lynch-sethuraman:1987,mogulskii:1993,de-acosta:1994,feng-kurtz:2006,budhiraja-dupuis-maroulas:2011}, with the quadratic action replaced by the Legendre transform of the noise CGF, similar to the WKB theory of jump processes in chemical kinetics~\cite{dykman-mori-ross-etal:1994}. This setting has recently been studied by Baule and Sollich~\cite{baule-sollich:2023} for one-dimensional escape problems (see also Refs.~\cite{dykman:2010,assaf-meerson:2017}). Among other results, they found that the escape rate is exponentially enhanced compared to Gaussian noise of the same variance, and that the instanton can contain jumps, depending on how fast the tail of the jump size distribution decays. We will revisit and interpret their results from the OC perspective.

The aim of our paper is to extend the OC formulation and prefactor computation to this class of L{\'e}vy-driven SDEs in arbitrary dimension, and to treat discontinuous instantons within the same framework. After reviewing the LDT setting and the three noise types used throughout in Sec.~\ref{sec:ldt}, our main contributions are:

\begin{enumerate}
\item We formulate the instanton problem for SDEs driven by light-tailed L{\'e}vy noise as an OC problem, whose running cost is the Legendre transform of the noise CGF, and discuss its efficient numerical solution with automatic differentiation, e.g., in JAX~\cite{bradbury-frostig-hawkins-etal:2018} (Sec.~\ref{sec:control}).
\item To compute instantons that contain jumps, we introduce a time reparameterization (TR) in the spirit of gMAM~\cite{heymann-vanden-eijnden:2008}, in which a jump corresponds to a finite displacement while physical time is frozen (Sec.~\ref{sec:time-reparam}).
\item We distinguish between jumps of the instanton that are merely admissible, i.e., have finite cost, and those that are optimal. Admissibility is decided by the recession function of the cost, whereas optimality is decided by its subleading behavior and drift. In particular, for the three classes of jump distributions of Baule and Sollich~\cite{baule-sollich:2023} and in accordance with their results, jumps are excluded for super-exponential and purely exponential tails, while they can be optimal if the exponential tail has an additional integrable algebraic decay (Sec.~\ref{sec:jumps}).
\item For continuous instantons, we derive the leading-order prefactor of probability density functions (PDFs) and tail probabilities via Riccati equations and Fredholm determinants. Compared to the Gaussian case, the only modification is a pointwise reweighting of the second variation by the square root of the covariance of the exponentially tilted noise along the instanton. We evaluate the prefactor analytically for one-dimensional gradient systems at long times, showing that it generically depends on the endpoint, and discuss partial results for the prefactor of jumping instantons (Sec.~\ref{sec:prefactor}).
\item We illustrate the approach on three examples (Sec.~\ref{sec:examples}): a two-dimensional nonlinear SDE~\cite{schorlepp-grafke-grauer:2021,schorlepp-tong-grafke-etal:2023}, a run-and-tumble particle (RTP) in an anisotropic trap with non-Gaussian orientational noise, and the Ghil--Sellers energy balance model (EBM) with L{\'e}vy-perturbed solar forcing~\cite{sellers:1969,ghil:1976,bodai-lucarini-lunkeit-etal:2015,lucarini-serdukova-margazoglou:2022}. Instantons and, where available, sharp estimates are compared against
Monte Carlo results.
\end{enumerate}
Figure~\ref{fig:concept} summarizes the main ideas for the two-dimensional example of Sec.~\ref{sec:2d}. Throughout, we consider instantons over a finite time horizon $[0,T]$. For the transitions between climate states in Sec.~\ref{sec:gsebm}, this approximates an infinite time horizon. A systematic extension to infinite time horizons, relevant for transition rates between metastable states, is a natural next step that we comment on in Sec.~\ref{sec:conclusion}. Our presentation is formal rather than mathematically rigorous. In particular, we derive the sharp asymptotic expansions only formally, and validate them numerically.

\paragraph*{Related work.} Closest to this work, Wei et al.~\cite{wei-gao-chen-etal:2022} formulated an OC problem for most likely transition paths of SDEs with super-exponentially light jumps in the same scaling, see also Ref.~\cite{li-duan-liu-etal:2020} for the corresponding Hamiltonian formalism. In Ref.~\cite{wei-gao-chen-etal:2022}, the control is the density of the controlled jump intensity with respect to the L{\'e}vy measure, i.e., a function of time and jump size, which is parameterized together with the path by neural networks. However, minimizing the cost of such controls subject to a prescribed mean forcing $\eta$ yields exactly the Legendre transform of the noise CGF, such that this lifting is unnecessary. Instead, it suffices to optimize over the noise realization $\eta(t) \in \RR^m$, as we do here. Hamiltonian-based minimum action methods, such as the recent method of Zakine and Vanden-Eijnden~\cite{zakine-vanden-eijnden:2023}, also apply to the LDT Hamiltonians considered here, but assume them to be twice differentiable, and hence finite, for all momenta. This excludes CGFs with bounded domain and discontinuous instantons from existing formulations.

%%%%%%%%%%%%%%%%%%%%%%%%%%%%%%%%%%%%%%%%%%%%%%%%%%%%%%%%%%%%%%%%%%%%%%%%%%%
\section{Large deviations for SDEs with light-tailed L{\'e}vy noise}
\label{sec:ldt}
%%%%%%%%%%%%%%%%%%%%%%%%%%%%%%%%%%%%%%%%%%%%%%%%%%%%%%%%%%%%%%%%%%%%%%%%%%%

\subsection{Setup and scaling limit}

Let $(Z_t)_{t \geq 0}$ be a L{\'e}vy process in $\RR^m$ with zero mean, Gaussian covariance $D \in \RR^{m \times m}$ and L{\'e}vy measure $\nu$, i.e., with CGF
\begin{align}
\Psi(p) &:= \frac{1}{t} \log \EE \left[ e^{\langle p, Z_t \rangle}\right]\,, \quad p \in \mathbb{R}^m \nonumber\\
&= \frac{1}{2} \langle p, D p \rangle + \int_{\RR^m} \left(e^{\langle p, y \rangle} - 1 - \langle p, y \rangle \right) \nu(\dd y)\,.
\label{eq:cgf}
\end{align}
We call the noise \emph{light-tailed} if $\Psi$ is finite on a neighborhood of the origin. Then, $\dom \Psi = \{p \in \RR^m \colon \Psi(p) < \infty\}$ is a convex set with $0$ in its interior, $\Psi$ is convex and smooth in the interior of its domain, and the covariance of $Z_1$ is $\nabla^2 \Psi(0) = D + \int y y^\top \nu(\dd y)$. For a small parameter $\eps > 0$,  we consider the rescaled L{\'e}vy process
\begin{align}
Z^\eps_t := \eps Z_{t / \eps}\,,
\label{eq:levy-scaling}
\end{align}
which has Gaussian covariance $\eps D$, while its jumps are of size $O(\eps)$ and occur with intensity $O(\eps^{-1})$. By construction,
\begin{align}
\log \EE \left[ \exp \left\{ \eps^{-1} \langle p, Z^\eps_t \rangle \right\} \right] = \eps^{-1} t \, \Psi(p)\,,
\label{eq:scaled-cgf}
\end{align}
and the covariance of $Z^\eps_t$ is $\eps t \nabla^2\Psi(0)$. A Gaussian process with the same covariance would be given by $\sqrt{\eps}\, \nabla^2\Psi(0)^{1/2} W_t$ with $(W_t)$ standard Brownian motion. For pure Brownian noise, Eq.~\eqref{eq:levy-scaling} is just the familiar diffusive scaling $\eps W_{t/\eps} \overset{d}{=} \sqrt{\eps} W_t$. For compound Poisson noise with jump rate $r_0$ and jump amplitude scale $a_0$, this corresponds to the scaling $r_0 = r/\eps$, $a_0 = \eps a$ considered in Ref.~\cite{baule-sollich:2023}, with variance per unit time of $r_0 a_0^2 = \eps r a^2$. Note that the scaling~\eqref{eq:levy-scaling} differs from $\eps Z_t$ (i.e., small jumps with fixed intensity), for which rare events are caused by a single or a few large jumps, depending on the tails, and exit times are not of Freidlin--Wentzell type~\cite{imkeller-pavlyukevich-wetzel:2009,imkeller-pavlyukevich-wetzel:2010}.

We consider the SDE
\begin{align}
\dd X^\eps_t = b(X^\eps_t) \, \dd t + \sigma(X_{t^-}^\eps) \, \dd Z^\eps_t\,, \quad X_0^\eps = x_0 \in \RR^n\,,
\label{eq:sde}
\end{align}
with drift $b \colon \RR^n \to \RR^n$ and diffusion coefficient matrix $\sigma \colon \RR^n \to \RR^{n \times m}$, which is allowed to be degenerate. Here, $X^\eps_{t^-} := \lim_{s \uparrow t} X^\eps_s$ denotes the state immediately before a jump at time $t$, while $X^\eps_t$ denotes the state after the jump. Evaluating $\sigma$ at the pre-jump state is the It{\^o} interpretation. Most of our examples have additive noise, $\sigma = \text{const.}$, but the EBM in Sec.~\ref{sec:gsebm} has multiplicative noise where the interpretation of the SDE~\eqref{eq:sde} matters. For multiplicative noise, the It{\^o} and Marcus interpretations (cf.\ Appendix~\ref{app:ito-marcus}, and Refs.~\cite{marcus:1981,kurtz-pardoux-protter:1995,applebaum:2009,chechkin-pavlyukevich:2014}) yield the same rate function if the instanton does not
jump, since the interpretations differ only by an $O(\eps)$ correction in that case.

We are interested in the probability of rare events of the form $\{f(X^\eps_T) \geq z\}$, or in the PDF $\rho_\eps(z)$ of $f(X^\eps_T)$ at~$z$, for a scalar observable $f \colon \RR^n \to \RR$ and a fixed final time $T > 0$, as $\eps \downarrow 0$. This is the setting in the two-dimensional example of Sec.~\ref{sec:2d}, and we focus on it here for clarity. Vector-valued observables, as for the RTP in Sec.~\ref{sec:rtp}, are treated in Appendix~\ref{app:multi-constraint}, and transitions between metastable states are discussed for the EBM in Sec.~\ref{sec:gsebm}.

\subsection{Sample path large deviations and the LDT Hamiltonian}

Let $L = \Psi^*$ denote the Legendre transform of the noise CGF,
\begin{align}
L(\eta) = \sup_{p \in \RR^m} \left\{ \langle p, \eta \rangle - \Psi(p) \right\}\,, \quad \eta \in \RR^m\,.
\label{eq:legendre}
\end{align}
Then, formally, the probability that the noise $\dot{Z}^\eps$ is close to a given realization $\eta \colon [0,T] \to \RR^m$ satisfies $\PP[\dot{Z}^\eps \approx \eta] \asymp \exp\{-\eps^{-1} \int_0^T L(\eta(t)) \, \dd t\}$. Minimizing this noise cost over controls that generate a prescribed absolutely continuous state path $\phi$ suggests the action
\begin{align}
 S[\phi] = \inf_\eta \textstyle \int_0^T L(\eta(t)) \, \dd t \; \text{s.t.} \; \begin{cases}
\dot{\phi} = b(\phi) + \sigma(\phi) \eta\,,\\
\phi(0) = x_0\,.
\end{cases}
\label{eq:action}
\end{align}
For $\sigma = \Id$, this is $S[\phi] = \int_0^T L(\dot{\phi} - b(\phi)) \, \dd t$, which reduces to the Freidlin--Wentzell action $\tfrac12 \int_0^T \langle \dot{\phi} - b(\phi), D^{-1}(\dot{\phi} - b(\phi)) \rangle \, \dd t$ for Gaussian noise with positive definite covariance $D$. Rigorous versions of this LDP for processes with independent increments and SDEs driven by Poisson random measures in the present scaling have been established in the literature~\cite{lynch-sethuraman:1987,mogulskii:1993,de-acosta:1994,leonard:2000,feng-kurtz:2006,budhiraja-dupuis-maroulas:2011,budhiraja-chen-dupuis:2013,gomes-hoegele:2021}.

An important subtlety, which will be central below, arises when $\dom \Psi$ is bounded. Then $L$ grows only linearly at infinity, and the action~\eqref{eq:action} has to be extended from absolutely continuous paths to paths of bounded variation that may contain jumps~\cite{lynch-sethuraman:1987,mogulskii:1993,leonard:2000}. The cost of a jump $v \in \RR^m$ of the integrated noise, which results in a jump $\Delta \phi = \sigma v$ of the path for additive noise, is given by the ``recession function''~\cite{rockafellar:1970}
\begin{align}
L_\infty(v) := \lim_{\kappa \downarrow 0} \kappa L(v / \kappa) = \sup_{p \in \dom \Psi} \langle p, v \rangle\,,
\label{eq:recession}
\end{align}
which is the support function of $\dom \Psi$ and is finite if and only if $\dom \Psi$ is bounded in direction $v$. For additive noise, the extended action combines the continuous contribution in Eq.~\eqref{eq:action} with the sum of the recession costs of the noise jumps, minimized over noise realizations that generate the state path.

The rate function of the observable $f(X^\eps_T)$ is obtained by contraction:
\begin{align}
I(z) = \inf \left\{ S[\phi] \colon \phi(0) = x_0\,, \; f(\phi(T)) = z \right\}\,,
\label{eq:rate-function}
\end{align}
such that $\PP[f(X^\eps_T) \geq z] \asymp \exp\{-I(z)/\eps\}$ for $z$ larger than the deterministic value $f(\bar{x}(T))$, where $\bar{x}$ solves $\dot{\bar{x}} = b(\bar{x})$, $\bar{x}(0) = x_0$. The minimizer $\phi_z$ is called ``instanton'' (while, by abuse of terminology, we also sometimes refer to the optimal control $\eta_z$ as an instanton in this paper). If $\phi_z$ is smooth and $\sigma(\phi_z)^\top p_z$ lies in the interior of $\dom \Psi$, it satisfies Hamilton's equations with the LDT Hamiltonian
\begin{align}
H(x,p) = \langle b(x), p \rangle + \Psi(\sigma(x)^\top p)\,,
\label{eq:hamiltonian}
\end{align}
namely
\begin{align}
\begin{cases}
\dot{\phi}_z = \nabla_p H(\phi_z, p_z) = b(\phi_z) + \sigma(\phi_z) \nabla \Psi(\sigma(\phi_z)^\top p_z)\,,\\
\dot{p}_z = -\nabla_x H(\phi_z, p_z)\,,\\
\phi_z(0) = x_0\,, \quad p_z(T) = \lambda_z \nabla f(\phi_z(T))\,,
\end{cases}
\label{eq:hamilton}
\end{align}
where the Lagrange multiplier $\lambda_z$ enforces $f(\phi_z(T)) = z$ and satisfies $\lambda_z = I'(z)$ wherever $I$ is differentiable. The optimal noise realization is $\eta_z = \nabla \Psi(\sigma^\top p_z)$, with inverse relation $\sigma^\top p_z = \nabla L(\eta_z)$, and the Hamiltonian is conserved, $H(\phi_z(t), p_z(t)) \equiv E_z$. Note that, in contrast to the Gaussian case, the momentum is restricted to $\sigma^\top p_z \in \dom \Psi$. At the boundary of this domain, the Hamiltonian description may fail and jumps can occur~\cite{baule-sollich:2023}. If the CGF has a finite limiting gradient at the boundary, continuous saturated arcs are also possible, see Sec.~\ref{sec:jumps}.

\subsection{Noise types}
\label{sec:noise-types}

\begin{table*}
\caption{Examples of the three noise types~\cite{baule-sollich:2023}, with unit variance per unit time, $\psi''(0) = 1$. For type C with $\alpha \in (1,2)$, the Legendre transform is exactly affine beyond the saturation threshold $\psi_\alpha'(1) = 2^{\alpha-2}/(\alpha-1)$, see Eq.~\eqref{eq:legendre-asymptotics}.}
\label{tab:noise-types}
\begin{ruledtabular}
\begin{tabular}{lllll}
type & L{\'e}vy measure $\omega(\dd y)$ & CGF $\psi(u)$ & $\dom \psi$ & $\psi^*(\eta)$ as $|\eta| \to \infty$\\
\hline
Gaussian & --- & $u^2/2$ & $\RR$ & $\eta^2 / 2$\\
A & $\tfrac12 \left(\delta_1(\dd y) + \delta_{-1}(\dd y)\right)$ & $\cosh u - 1$ & $\RR$ & $|\eta| (\log(2|\eta|) - 1) + 1 + o(1)$\\
B & $\tfrac14 e^{-|y|} \, \dd y$ & $\frac{u^2}{2(1-u^2)}$ & $(-1,1)$ & $|\eta| - \sqrt{|\eta|} + \tfrac38 + o(1)$\\
C & $\frac{|y|^{-1-\alpha}e^{-|y|}}{2 \Gamma(2-\alpha)} \, \dd y$ & $\frac{(1+u)^\alpha + (1-u)^\alpha - 2}{2 \alpha (\alpha-1)}$ & $[-1,1]$ & $|\eta| - c_\alpha + o(1)$ for $\alpha \in (0,1)$\\
 & & & & $|\eta| - c_\alpha$ for $\alpha \in (1,2)$, $|\eta| \geq \psi_\alpha'(1)$
\end{tabular}
\end{ruledtabular}
\end{table*}

\begin{figure*}
\centering
\includegraphics[width=\textwidth]{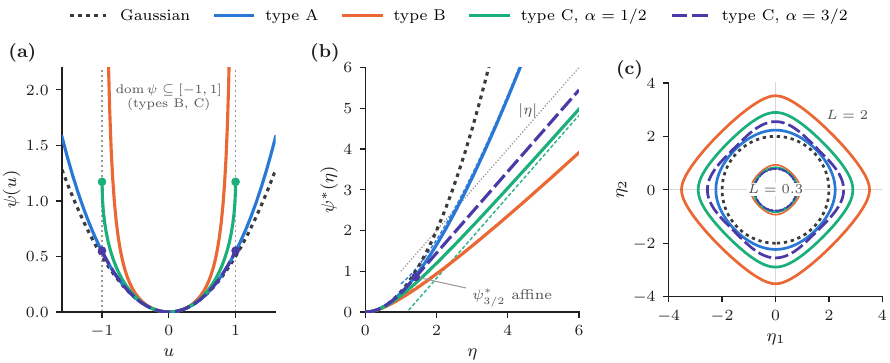}
\caption{The noise types of Table~\ref{tab:noise-types}, compared to Gaussian noise of the same variance. (a)~CGFs $\psi(u)$; for types B and C, $\dom \psi$ is bounded. (b)~Legendre transforms $\psi^*(\eta)$, with the large-$|\eta|$ asymptotics~\eqref{eq:legendre-asymptotics} as thin dashed lines. (c)~Level sets of the cost $L(\eta) = \psi^*(\eta_1) + \psi^*(\eta_2)$ of two independent noise components. Small costs probe only the covariance, while level sets of large costs approach the diamonds of an $\ell^1$ norm for types B and C.}
\label{fig:cgf}
\end{figure*}

For the remainder of this paper, we take the noise components to be independent, $\Psi(p) = \sum_{i=1}^m \Psi_i(p_i)$ and hence $L(\eta) = \sum_{i=1}^m L_i(\eta_i)$, and each non-Gaussian component to be a symmetric pure jump process with intensity scale $r_i > 0$ and jump amplitude scale $a_i > 0$,
\begin{align}
\Psi_i(p_i) &= r_i \psi(a_i p_i)\,, \quad L_i(\eta_i) = r_i \psi^*\!\left(\frac{\eta_i}{r_i a_i}\right)\,, \nonumber\\
\psi(u) &= \int_\RR \left(e^{u y} - 1 - u y \right) \omega(\dd y)\,.
\label{eq:scalar-noise}
\end{align}
Here, $\omega$ is a symmetric L{\'e}vy measure normalized to unit second moment, $\psi''(0) = \int y^2 \omega(\dd y) = 1$, such that the variance per unit time of component $i$ is $r_i a_i^2$. When $\omega$ has a density $\omega(y)$, the L{\'e}vy measure of component $i$ is $\nu_i(\dd y) = (r_i/a_i) \omega(y/a_i) \, \dd y$. Thus, $r_i$ is the total jump rate only if $\omega$ has unit total mass; in general, the rate is $r_i \omega(\RR)$, which may be infinite. Following Ref.~\cite{baule-sollich:2023}, we distinguish three types of jump distributions, see Table~\ref{tab:noise-types} and Fig.~\ref{fig:cgf}:
\begin{itemize}
\item \textbf{Type A}: $\dom \psi = \RR$, i.e., jump distributions with super-exponential tails. Our concrete example of this, as in Ref.~\cite{baule-sollich:2023}, are jumps of size $\pm 1$, with baseline measure $\omega = (\delta_1 + \delta_{-1})/2$, where $\delta_c$ denotes a unit point mass at $c$, and $\psi(u) = \cosh u - 1$. The Legendre transform grows superlinearly, $\psi^*(\eta) \sim |\eta| \log |\eta|$, but slower than for Gaussian noise.
\item \textbf{Type B}: $\dom \psi$ is bounded and $\psi$ blows up at its boundary. Our example has Laplace-distributed jumps, with baseline L{\'e}vy density $\omega(y) = \tfrac14 e^{-|y|}$ (with total mass $1/2$ as in Ref.~\cite{baule-sollich:2023}, so the actual rate will be $r/2$ in this case), and $\psi(u) = u^2 / (2(1-u^2))$ on $\dom \psi = (-1,1)$.
\item \textbf{Type C}: $\dom \psi$ is bounded and $\psi$ remains finite on its boundary. Our example is an infinite-activity process with baseline L{\'e}vy density $\omega(y) = |y|^{-1-\alpha} e^{-|y|} / (2 \Gamma(2 - \alpha))$ with $\alpha \in (0,2)$, $\alpha \neq 1$, for which $\dom \psi = [-1,1]$ and $\psi_\alpha(u) = [(1+u)^\alpha + (1-u)^\alpha - 2] / [2 \alpha (\alpha - 1)]$. We use $\alpha = 1/2$ and $\alpha = 3/2$ below for numerical experiments.
\end{itemize}
More generally, for jump densities $\omega(y) = c(y) e^{-|y|}$ with subexponential $c$, $\psi$ is of type C if $\int^\infty c(y) \, \dd y < \infty$, and of type B otherwise~\cite{baule-sollich:2023}. The singularity of the type C example at the origin makes it infinite-activity, but the classification above depends only on the tails. We use these three examples throughout the paper and refer to them simply as type A, B and C noise. Note that, for any symmetric baseline L{\'e}vy measure with unit second moment, $\cosh v - 1 \geq v^2/2$ implies $\psi(u) \geq u^2/2$, and hence $\psi^*(\eta) \leq \eta^2/2$. Thus, at equal variance, the cost of every noise realization, and therefore every rate function, is bounded from above by its Gaussian counterpart~\cite{baule-sollich:2023}.

For types B and C, the recession function~\eqref{eq:recession} of $L_i$ is $L_{i,\infty}(v) = |v| / a_i$. In other words, the large-noise cost approaches a weighted $\ell^1$ norm, whereas the Gaussian cost is quadratic (see Fig.~\ref{fig:cgf}(c)). The subleading behavior of type B and C differs, however, and this difference is decisive for whether instantons jump, see Sec.~\ref{sec:jumps}. Straightforward calculations show, as $|\eta| \to \infty$, that
\begin{align}
\begin{cases}
\psi^*_{\rm B}(\eta) &= |\eta| - \sqrt{|\eta|} + \tfrac38 + o(1)\,,\\
\psi^*_\alpha(\eta) &= |\eta| - c_\alpha + d_\alpha |\eta|^{-\gamma_\alpha} + o(|\eta|^{-\gamma_\alpha})\,,\\
&\qquad \alpha \in (0,1)\,,\\
\psi^*_\alpha(\eta) &= |\eta| - c_\alpha\,, \qquad \alpha \in (1,2)\,,\\
&\qquad \text{exactly for } |\eta| \geq \psi_\alpha'(1) = \frac{2^{\alpha-2}}{\alpha - 1}\,,
\end{cases}
\label{eq:legendre-asymptotics}
\end{align}
with $c_\alpha := \psi_\alpha(1) = (2^{\alpha - 1} - 1) / (\alpha (\alpha - 1)) > 0$, $\gamma_\alpha = \alpha / (1 - \alpha)$, and $d_\alpha = (2 \alpha)^{-1} (2 (1-\alpha))^{-\gamma_\alpha}$.
For $\alpha \in (1,2)$, the derivative $\psi_\alpha'$ is bounded on $[-1,1]$, such that the Legendre maximizer reaches the boundary $u = \pm 1$ at finite $\eta$, beyond which $\psi_\alpha^*$ is exactly affine and $(\psi^*_\alpha)'' = 0$. We refer to this regime as \emph{saturation}.

%%%%%%%%%%%%%%%%%%%%%%%%%%%%%%%%%%%%%%%%%%%%%%%%%%%%%%%%%%%%%%%%%%%%%%%%%%%
\section{Instantons via optimal control}
\label{sec:control}
%%%%%%%%%%%%%%%%%%%%%%%%%%%%%%%%%%%%%%%%%%%%%%%%%%%%%%%%%%%%%%%%%%%%%%%%%%%

\subsection{Control formulation}
\label{sec:control-formulation}

Equations~\eqref{eq:action} and~\eqref{eq:rate-function} immediately suggest viewing the noise realization $\eta$ as a control. Denoting by $\phi^\eta$ the solution of the controlled system $\dot{\phi}^\eta = b(\phi^\eta) + \sigma(\phi^\eta) \eta$, $\phi^\eta(0) = x_0$, and by $F[\eta] := f(\phi^\eta(T))$ the observable as a functional of the control, the rate function is
\begin{align}
I(z) = \inf_{\eta \colon F[\eta] = z} \int_0^T L(\eta(t)) \, \dd t\,.
\label{eq:control-problem}
\end{align}
If the instanton contains a jump, the infimum need not be attained by an ordinary time-dependent control. We address this case in Sec.~\ref{sec:time-reparam} and assume a smooth, nonsaturated minimizer $\eta_z$ for now. Problem~\eqref{eq:control-problem} is formally identical to the Gaussian case~\cite{schorlepp-tong-grafke-etal:2023,schorlepp-grafke:2025}, except that the quadratic running cost $\frac12 \norm{\eta}^2$ is replaced by $L$. With $\phi_z := \phi^{\eta_z}$, the first-order optimality condition of the constrained optimization problem~\eqref{eq:control-problem} reads
\begin{align}
\nabla L(\eta_z(t)) = \lambda_z \left.\frac{\delta F}{\delta \eta(t)}\right|_{\eta_z} = \sigma(\phi_z(t))^\top p_z(t)\,,
\label{eq:first-order}
\end{align}
where $p_z$ solves the adjoint equation $\dot{p}_z = -[\nabla b(\phi_z) + \nabla \sigma(\phi_z) \eta_z]^\top p_z$ with $p_z(T) = \lambda_z \nabla f(\phi_z(T))$. In the nonsaturated regime, where $\sigma^\top p_z$ lies in the interior of $\dom \Psi$ and $\nabla^2 \Psi$ is positive definite, $\nabla L$ and $\nabla \Psi$ are mutually inverse by Legendre duality. Thus, Eq.~\eqref{eq:first-order} is equivalent to $\eta_z = \nabla \Psi(\sigma^\top p_z)$, and the adjoint variable is precisely the momentum in Hamilton's equations~\eqref{eq:hamilton}. The OC formulation thus contains the same information as the Hamiltonian formulation, but it is more convenient in practice for the same reasons as in the Gaussian case, as outlined in Sec.~\ref{sec:intro}.

\paragraph*{Numerical solution.}
We follow a discretize-then-optimize approach. On a uniform time grid $t_k = k \Delta t$, the controlled dynamics is discretized with an explicit scheme (forward Euler in the simplest case), the action by a Riemann sum $\Delta t \sum_k L(\eta_k)$, and the terminal constraint $F_N(\eta) = z$ for the discretized noise-to-observable map $F_N$ is enforced through an augmented Lagrangian method (which does not require convexity of $z \mapsto I(z)$~\cite{alqahtani-grafke:2021,schorlepp-grafke:2025}), with each inner unconstrained problem solved by L-BFGS. Discretizing first ensures that action, multiplier and second variation all refer to the same discrete functional, which is important for optimization as well as consistent prefactor computations in Sec.~\ref{sec:prefactor}. The control optimizations use JAX~\cite{bradbury-frostig-hawkins-etal:2018,schorlepp-grafke:2025} for automatic differentiation, providing gradients and Hessian-vector products through the time stepping, as well as just-in-time compilation (and GPU support, if desired).

For Gaussian and type A noise, the cost function $L$ is available in closed form. For types B and C, we evaluate $L_i(\eta_i)$ by solving the scalar equation $\eta_i = \Psi_i'(u_i)$ for $u_i$ in the interior of $\dom \Psi_i$ with a projected Newton iteration whose updates are clipped to remain in the interior of the domain. Here, $u$ is the noise-space dual variable, equal to $\sigma^\top p_z$ along the instanton. In the nonsaturated regime, where $\nabla^2 \Psi(u)$ is positive definite, we supply derivatives through the implicit relations
\begin{align}
\nabla L(\eta) = u\,, \qquad \nabla^2 L(\eta) = \left[\nabla^2 \Psi(u)\right]^{-1}\,,
\label{eq:implicit-legendre}
\end{align}
instead of differentiating through the Newton iterations. For type C with $\alpha \in (1,2)$ and $|\eta_i| \geq r_i a_i \psi_\alpha'(1)$, this equation has no solution in the interior, and we set $u_i = \operatorname{sgn}(\eta_i)/a_i$ on the boundary, with $L_i$ affine and $L_i'' = 0$. Near the boundary of $\dom \Psi$, the large derivatives of $\Psi$ make numerical evaluation delicate, while the small curvature of $L$ makes the optimization poorly conditioned. Convergence is therefore typically slower for types B and C than for Gaussian and type A noise. Since the optimization problems we consider are nonconvex due to the nonlinear constraint, we further check that independent initializations, e.g., with the Gaussian instanton, lead to the same minimizer. For problems with additional structure, a reduction of the number of optimization variables can improve robustness, see Sec.~\ref{sec:rtp}.

\subsection{Resolving jumps via time reparameterization}
\label{sec:time-reparam}

If $\dom \Psi$ is bounded, the instanton may contain jumps or very fast transitions, which cannot be resolved on a uniform time grid with finite controls. Following the idea of gMAM~\cite{heymann-vanden-eijnden:2008} -- while, in a certain sense, reversing its logic -- we introduce an artificial time $s \in [0,1]$ and a nonnegative time density $\kappa(s) = \dd t / \dd s$, with physical time $\tau(s) = \int_0^s \kappa(s') \, \dd s'$ and $\tau(1) = T$. We represent the physical path by an absolutely continuous pair $(\tau(s),q(s))$: on continuous portions, $q(s) = \phi(\tau(s))$, while each jump is resolved by an inserted transition from its pre-jump to its post-jump state at fixed~$\tau$. Wherever $\kappa > 0$, the dynamics becomes $q' = \kappa [b(q) + \sigma(q) \eta(\tau(s))]$. To allow finite state motion when $\kappa$ vanishes, we introduce a rescaled control $\hat{\eta}$, equal to $\kappa \eta(\tau(s))$ where $\kappa > 0$ and defined independently on frozen-time intervals. This gives the time-reparameterized (TR) control problem
\begin{align}
&I(z) = \min_{\hat{\eta}, \kappa} \int_0^1 \bar{L}(\hat{\eta}(s), \kappa(s)) \, \dd s \label{eq:tr-problem}\\
&\text{s.t.} \quad
\begin{cases}
q' = \kappa b(q) + \sigma(q) \hat{\eta}\,, \; q(0) = x_0\,, \; f(q(1)) = z\,,\\
\kappa(s) \geq 0\,, \quad \int_0^1 \kappa(s) \, \dd s = T\,,
\end{cases}\nonumber
\end{align}
where the running cost is the ``closed perspective function'' of $L$~\cite{rockafellar:1970},
\begin{align}
\bar{L}(\hat{\eta}, \kappa) =
\begin{cases}
\kappa L(\hat{\eta} / \kappa)\,, & \kappa > 0\,,\\
L_\infty(\hat{\eta})\,, & \kappa = 0\,.
\end{cases}
\label{eq:perspective}
\end{align}
The perspective function is jointly convex in $(\hat{\eta}, \kappa)$, and its closure at $\kappa = 0$ is the recession function~\eqref{eq:recession}. A jump of the instanton in physical time is now represented by an interval of artificial time on which $\kappa = 0$ while $q' \neq 0$, i.e., the system moves while physical time is frozen. For superlinear $L$, such intervals have infinite cost and Eq.~\eqref{eq:tr-problem} reduces to Eq.~\eqref{eq:control-problem}. For multiplicative noise, the state follows the flow $q' = \sigma(q) \hat{\eta}$ during a jump, as in the Marcus interpretation of Eq.~\eqref{eq:sde}~\cite{marcus:1981,kurtz-pardoux-protter:1995}.
Hence, for jumping instantons with multiplicative noise, Eq.~\eqref{eq:tr-problem} yields the rate function of the Marcus SDE, cf.\ Sec.~\ref{sec:gsebm}.

As in gMAM, problem~\eqref{eq:tr-problem} is invariant under further reparameterizations of artificial time. Minimizers are therefore never isolated, but come in families that differ only in how artificial time is spent, and only the physical graph $\{(\tau(s), q(s))\}$ is meaningful. gMAM removes this degeneracy by an arclength constraint~\cite{heymann-vanden-eijnden:2008}. In this paper, for simplicity, we deal with this issue through additional regularization terms added to the objective, as detailed in Appendix~\ref{app:regularize-details}. Note that the TR problem remains nearly degenerate, though, and its second variation is hence not suitable for prefactor computations. Instead, continuous instantons are computed, or refined, in the fixed-time formulation~\eqref{eq:control-problem} for this purpose. We further remark that as an alternative to our TR approach, jumps can also be built in explicitly, by optimizing over jump times and sizes together with the continuous controls between the jumps. When solving the corresponding Hamiltonian equations of motion~\eqref{eq:hamilton} along continuous segments, we refer to the resulting method as ``Hamiltonian shooting''.

\subsection{Admissible versus optimal jumps}
\label{sec:jumps}

For additive noise with $\sigma = \Id$, a jump $v$ of the instanton is \emph{admissible}, i.e., has finite cost, if and only if $L_\infty(v) < \infty$~\cite{lynch-sethuraman:1987,mogulskii:1993,leonard:2000}, which for types B and C holds in every direction. 
Whether a jump is also \emph{optimal} is a different question, whose answer depends on the subleading behavior of $L$ and on the drift. We first give a simple local argument based on Taylor expansion, and then state the precise optimality conditions and what they imply for jumps in the optimal path.

\begin{figure*}
\centering
\includegraphics[width=\textwidth]{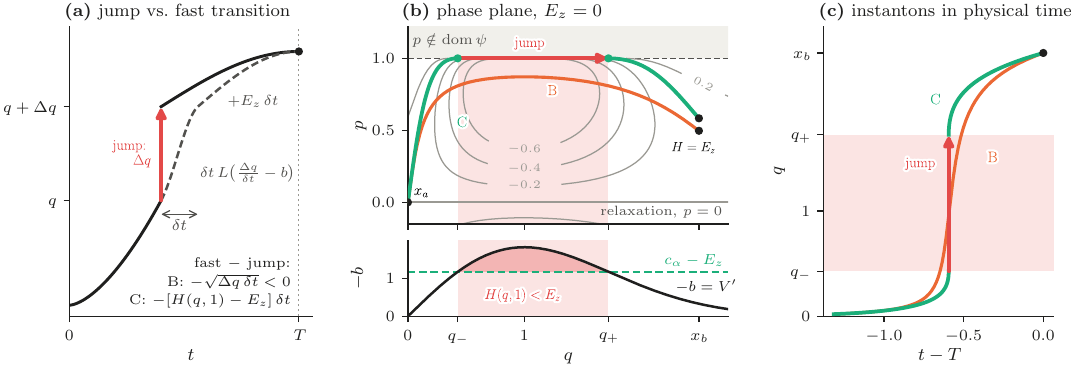}
\caption{Optimal jumps in one dimension. (a)~Sketch of the local argument~\eqref{eq:segment-vs-jump}: cost of a piece of a jump (red) and of a fast continuous transition (dashed). (b)~Instantons in the phase plane (thick lines) on the level set $\{H = E_z\}$. Thin gray lines indicate other level sets of $H$ for type C. The type C instanton jumps along the boundary $p = 1$ of $\dom \psi$ across the interval $[q_-, q_+]$ (shaded) in which the jump criterion~\eqref{eq:jump-criterion} holds (bottom). (c)~The same instantons in physical time. Panels (b) and (c) use $b = -V'$ with $V'(q) = 3 q e^{-q^2/2}$, $r = a = 1$ and $E_z = 0$, with type B (orange) and type C noise ($\alpha = 1/2$, green).}
\label{fig:jumps-1d}
\end{figure*}

\paragraph*{Local argument.}
Consider a scalar noisy coordinate $q(t) \in \mathbb{R}$ with additive noise, in units $r = a = 1$, and a jump of the instanton in the positive direction that passes through a point $q$. We compare a short piece $[q, q + \Delta q]$ of this jump, which costs $L_\infty(\Delta q) = \Delta q$, with a fast continuous transition by the same amount during a short time $\delta t \ll \Delta q$, see Fig.~\ref{fig:jumps-1d}(a). For small $\Delta q$, the drift is approximately constant, $b \approx b(q)$, during the transition, the required control is $\eta \approx \Delta q / \delta t - b(q)$, and the cost is $\delta t \, L(\Delta q / \delta t - b(q))$. Since $T$ is fixed, the time $\delta t$ is missing for the remainder of the path, which increases its action by $E_z \delta t$ to leading order, where $E_z$ is the conserved value of the Hamiltonian along the instanton (by the Hamilton--Jacobi relation $\partial_T I = -E_z$). Inserting the asymptotics~\eqref{eq:legendre-asymptotics}, the cost of the continuous transition exceeds that of the jump by
\begin{align}
&\delta t \, L\left(\frac{\Delta q}{\delta t} - b(q)\right) + E_z \delta t - \Delta q \nonumber\\
&=
\begin{cases}
-\sqrt{\Delta q \, \delta t} + O(\delta t)\,, & \text{type B}\,,\\
-\left[H(q, 1) - E_z\right] \delta t + o(\delta t)\,, & \text{type C}\,,
\end{cases}
\label{eq:segment-vs-jump}
\end{align}
as $\delta t \downarrow 0$, up to corrections of order $\Delta q \, \delta t$ from the variation of the drift, where $H(q, 1) = b(q) + \psi(1) = b(q) + c_\alpha$ is the Hamiltonian at the boundary momentum $p = 1$.
For type B, the subleading term $-\sqrt{|\eta|}$ of the cost makes the difference negative, irrespective of the drift. A fast continuous transition is hence always cheaper than a jump, and type B instantons never jump, although they can contain very fast continuous transitions, see Fig.~\ref{fig:jumps-1d}(c). For type C, the cost is affine up to $o(1)$, $L(\eta) = |\eta| - c_\alpha + o(1)$, and the sign of the difference is decided by the drift. Jumping through $q$ is favorable if
\begin{align}
H(q, 1) = b(q) + c_\alpha < E_z\,,
\label{eq:jump-criterion}
\end{align}
i.e., if the drift opposing the jump is sufficiently strong at $q$, where $c_\alpha$ is defined under Eq.~\eqref{eq:legendre-asymptotics}.

\paragraph*{Optimality conditions.}
This heuristic argument can be made precise, and extended to higher dimensions, through the first-order necessary conditions (Pontryagin's maximum principle~\cite{bryson-ho:1975}) for the TR problem~\eqref{eq:tr-problem}, which treats continuous parts and jumps of the instanton on the same footing. We consider additive noise, $\sigma = \Id$, and denote by $p(s)$ the adjoint variable of the state $q(s)$ and by $E_z$ the multiplier of the time constraint $\int_0^1 \kappa \, \dd s = T$. On continuous parts of the instanton, $p$ coincides with the momentum $p_z$ in Hamilton's equations~\eqref{eq:hamilton}, and $E_z$ with the conserved value of the Hamiltonian. The maximum principle states that, for almost every $s$, the controls $(\hat{\eta}(s), \kappa(s))$ maximize
\begin{align*}
h(\hat{\eta}, \kappa) := \langle p, \kappa b(q) + \hat{\eta} \rangle - \bar{L}(\hat{\eta}, \kappa) - E_z \kappa\,.
\end{align*}
Since $\bar{L}$ is the perspective function of $L$, we have $\sup_{\hat{\eta}} \{ \langle p, \hat{\eta} \rangle - \kappa L(\hat{\eta}/\kappa) \} = \kappa \sup_{\eta} \{ \langle p, \eta \rangle - L(\eta) \} = \kappa \Psi(p)$ for $\kappa > 0$ by Legendre duality, such that the maximum of $h$ over $\hat{\eta}$ at fixed $\kappa > 0$ is $\kappa [H(q,p) - E_z]$. This has two consequences. First, maximizing over $\kappa \geq 0$ requires $H(q,p) \leq E_z$ everywhere, with equality wherever $\kappa > 0$, i.e., when physical time runs. Second, where $\kappa = 0$, the state can still move, $q' = \hat{\eta} \neq 0$, but only if such a nonzero $\hat{\eta}$ maximizes $\langle p, \hat{\eta} \rangle - L_\infty(\hat{\eta})$. Since $L_\infty$ is the support function of $\dom \Psi$, this requires $p$ to lie on the boundary of $\dom \Psi$, and $\hat{\eta}$ to point in the direction of an outer normal of $\dom \Psi$ at $p$. Moreover, the adjoint equation $p' = -\kappa \nabla b(q)^\top p$ implies that $p$ is constant while $\kappa = 0$. Hence, along a jump from $q_-$ to $q_+$,
\begin{enumerate}
\item[(i)] the momentum $p_z$ is constant and lies on the boundary of $\dom \Psi$; for independent noise components, only components with $|p_{z,i}| = 1/a_i$ can jump, in the direction of $\operatorname{sgn} p_{z,i}$;
\item[(ii)] $H(q, p_z) \leq E_z$ for all $q$ along the path of the jump, with equality at its endpoints $q_\pm$ if these connect to continuous parts of the instanton.
\end{enumerate}
These are the Weierstrass--Erdmann corner conditions~\cite{bryson-ho:1975} of the present problem, and condition (ii) is the multidimensional version of Eq.~\eqref{eq:jump-criterion}, which has to hold along the whole jump, with equality at its endpoints. For type B, $\Psi \to \infty$ at the boundary of $\dom \Psi$, such that (ii) cannot be satisfied, and type B instantons do not jump in any dimension. For type C in one dimension, (i) and (ii) have a simple interpretation in the phase plane: the continuous parts of the instanton follow the level set $\{H = E_z\}$, and a jump is a horizontal segment on the line $p = 1/a$, which bridges exactly the interval in which this level set leaves $\dom \Psi$, i.e., in which Eq.~\eqref{eq:jump-criterion} holds. In the long-time limit $E_z = 0$ of escape problems with $b = -V'$, this is the region $\{V'(q) > r a \psi(1) \}$ identified in Ref.~\cite{baule-sollich:2023}. In higher dimensions, (i) and (ii) are necessary, but not sufficient, and whether the instanton actually jumps has to be decided, in general, by solving Eq.~\eqref{eq:tr-problem} numerically.

Figure~\ref{fig:jumps-1d} illustrates the one-dimensional picture in an example. We take an endpoint $x_b > q_+$ beyond the landing point of the type C jump described below. On its continuous portions, the instanton from the stable fixed point $x_a = 0$ follows the branch $\psi(p)/p = V'(q)$ of the level set $\{H = 0\}$ of $H(q,p) = -V'(q) p + \psi(p)$. Since $\psi$ is convex with $\psi(0) = 0$, the left-hand side increases in $p$. For type B, it diverges as $p \uparrow 1$, such that this branch exists for all $q$ and remains inside $\dom \psi = (-1,1)$. For type C, it is bounded by $\psi(1) = c_\alpha$, such that the branch reaches the boundary $p = 1$ at the point $q_-$ where $V'(q_-) = c_\alpha$, and returns only at $q_+ > q_-$, where $V'$ has decreased to $c_\alpha$ again, see Fig.~\ref{fig:jumps-1d}(b). In between, the instanton jumps at constant momentum $p = 1$. In physical time, the type C instanton thus jumps from $q_-$ to $q_+$, whereas the type B instanton crosses this interval quickly, but continuously, see Fig.~\ref{fig:jumps-1d}(c).

\paragraph*{Saturated arcs.}
For type C noise with $\alpha \in (1,2)$, a third possibility arises besides continuous parts with $\sigma^\top p_z$ in the interior of $\dom \Psi$ and jumps. Since $\psi_\alpha'$ is bounded, a component of the noise momentum $\sigma^\top p_z$ can reach the boundary of its CGF domain while the control remains finite, and stay there for some time interval. On such an arc, the corresponding scalar cost $L_i$ is affine and $(\nabla L(\eta_z))_i = (\sigma^\top p_z)_i$ is constant (the unsaturated components need not be constant). In the language of OC, this is a singular arc~\cite{bryson-ho:1975}. For additive noise with $\sigma = \Id$, saturation of component~$i$ requires $\frac{\dd}{\dd t} p_{z,i} = -\partial_{x_i} \langle b(\phi_z), p_z \rangle = 0$ on the whole interval. In one dimension, this is only possible if $b' = 0$ there, which is non-generic. Such arcs occur for the RTP in Sec.~\ref{sec:rtp}.

\paragraph*{Degenerate or multiplicative noise.} For Marcus SDEs, our reasoning carries over to multiplicative or degenerate $\sigma(x)$, mutatis mutandis. With $\hat{\eta}$ replaced by $\sigma(q)\hat{\eta}$ in $h$, the maximum principle still requires $H(q,p) \leq E_z$ and the noise momentum $\sigma(q)^\top p$ to lie on the boundary of $\dom \Psi$ during jumps. Hence, type B instantons still do not jump. The only changes are that jumps follow the flow $q' = \sigma(q)\hat{\eta}$ and that $p$ is no longer constant along a jump. For It{\^o} SDEs, the situation can be different, since a single large noise jump~$v$ then moves the state along the straight segment from~$q_-$ to $q_+ = q_- + \sigma(q_-) v$, at cost $L_\infty(v)$, with $\sigma$ frozen at the pre-jump state. If the noise amplitude decreases along the jump, this can be cheaper than any continuous path, such that It{\^o} instantons may jump even for type B noise.

%%%%%%%%%%%%%%%%%%%%%%%%%%%%%%%%%%%%%%%%%%%%%%%%%%%%%%%%%%%%%%%%%%%%%%%%%%%
\section{Sharp large deviation estimates: prefactors}
\label{sec:prefactor}
%%%%%%%%%%%%%%%%%%%%%%%%%%%%%%%%%%%%%%%%%%%%%%%%%%%%%%%%%%%%%%%%%%%%%%%%%%%

The rate function $I(z)$ determines probabilities only up to subexponential factors. For a quantitatively accurate estimate at finite $\eps$, we also need the leading-order prefactor,
\begin{align}
\begin{cases}
\rho_\eps(z) \overset{\eps \downarrow 0}{\sim} \frac{C(z)}{\sqrt{2 \pi \eps}} \exp\left\{-\frac{I(z)}{\eps}\right\}\,,\\
\PP\left[f(X^\eps_T) \geq z\right] \overset{\eps \downarrow 0}{\sim} \sqrt{\frac{\eps}{2 \pi}} \frac{C(z)}{\lambda_z} \exp\left\{-\frac{I(z)}{\eps}\right\}\,.
\end{cases}
\label{eq:sharp-asymptotics}
\end{align}
These expressions assume an isolated, non-degenerate instanton (contributions from several isolated minimizers must be summed). In this section, we take additive noise, and $\sigma = \Id$ for notational simplicity.
All instantons are parameterized by physical time on $[0,T]$, i.e., we work with the fixed-time problem~\eqref{eq:control-problem}.

\subsection{Continuous instantons: Fredholm determinant and Riccati formulation}
\label{sec:fredholm}

We give a formal derivation through an exponential change of measure, which is also the basis of instanton-based importance sampling~\cite{bucklew:2013,ebener-margazoglou-friedrich-etal:2019}, as used in Sec.~\ref{sec:rtp}.
For a deterministic, time-dependent tilt $p \colon [0,T] \to \RR^m$ in the interior of $\dom \Psi$, we define the path measure $\mathbb{Q}$ by
\begin{align}
\frac{\dd \mathbb{Q}}{\dd \PP} =  \textstyle\exp\left\{\frac{1}{\eps} \int_0^T \langle p(t), \dd Z^\eps_t \rangle - \frac{1}{\eps} \int_0^T \Psi(p(t)) \, \dd t \right\}\,,
\label{eq:tilt}
\end{align}
which has unit mean by Eq.~\eqref{eq:scaled-cgf}, since $Z^\eps$ has independent increments. Under $\mathbb{Q}$, the increments of $Z^\eps$ remain independent, with $\log \EE_{\mathbb{Q}}[\exp\{\eps^{-1} \langle k, \dd Z^\eps_t \rangle\}] = \eps^{-1} [\Psi(p(t) + k) - \Psi(p(t))] \, \dd t$. In other words, the Gaussian part of $Z^\eps$ acquires the drift $D p(t)$, and its L{\'e}vy measure $\nu_\eps$ is replaced by $e^{\langle p(t), y \rangle / \eps} \nu_\eps(\dd y)$, which favors jumps in the direction of the tilt. Consequently, $Z^\eps$ has mean $\nabla \Psi(p(t))$ and covariance $\eps \nabla^2 \Psi(p(t))$ per unit time under $\mathbb{Q}$. For the tilt $p = p_z$, the noise under $\mathbb{Q}$ is therefore centered at the instanton noise, $\dot{Z}^\eps = \eta_z + \sqrt{\eps} M_z \xi$, with $\xi$ Gaussian white noise to leading order in $\eps$ and
\begin{align}
M_z(t) := \left[\nabla^2 \Psi(p_z(t))\right]^{1/2}\,,
\label{eq:whitening}
\end{align}
since higher-order cumulants of $\xi$ under $\mathbb{Q}$ are suppressed by factors of $\sqrt{\eps}$.
A short calculation, documented in Appendix~\ref{sec:fredholm-tilt}, then yields Eq.~\eqref{eq:sharp-asymptotics} with prefactor
\begin{align}
C(z) &= \frac{|\lambda_z|}{\norm{M_z p_z}_{L^2} \, \det\left(\Id - A_z\right)^{1/2}}\,, \nonumber\\
A_z &:= \lambda_z \, \ppr_z M_z \frac{\delta^2 F}{\delta \eta^2}[\eta_z] M_z \ppr_z\,,
\label{eq:fredholm-prefactor}
\end{align}
where $\ppr_z$ denotes the $L^2$-orthogonal projection onto the complement of $M_z p_z$, which is proportional to the whitened gradient $M_z \, \delta F / \delta \eta$. For an alternative derivation of Eq.~\eqref{eq:fredholm-prefactor} via path integrals, see Appendix~\ref{sec:fredholm-PI}.

For Gaussian noise (and $\sigma = \operatorname{Id}$) with $\Psi(p) = \frac12 \norm{p}^2$, we have $M_z = \Id$ and $\norm{p_z}_{L^2} = \norm{\eta_z}_{L^2} = \sqrt{2 I(z)}$, and Eq.~\eqref{eq:fredholm-prefactor} reduces to the previously derived Gaussian prefactor~\cite{schorlepp-tong-grafke-etal:2023,schorlepp-grafke:2025}. For L{\'e}vy noise, the only modification is the pointwise multiplication of input and output of the second variation operator by $M_z(t)$, which is known along the instanton in closed form or via Eq.~\eqref{eq:implicit-legendre}. The operator $A_z$ is trace-class under the same conditions as in the Gaussian case, and its Fredholm determinant can be evaluated matrix-free through its dominant eigenvalues, computed by Hessian-vector products~\cite{schorlepp-tong-grafke-etal:2023,schorlepp-grafke:2025}. For several terminal constraints instead of a single, scalar one, an analogous expression holds, see Appendix~\ref{app:multi-constraint}.

The prefactor can equivalently be computed by integrating a forward matrix Riccati equation along the instanton. This formulation only involves the linearization of Hamilton's equations~\eqref{eq:hamilton} around the instanton, and hence carries over from the Gaussian case, with the LDT Hamiltonian~\eqref{eq:hamiltonian} in place of the Gaussian one~\cite{schorlepp-grafke-grauer:2023,grafke-schaefer-vanden-eijnden:2023}. Let $S(x,t)$ denote the minimal action to reach $x$ at time $t$ from $x_0$, such that $p_z(t) = \nabla_x S(\phi_z(t), t)$. Its inverse Hessian $Q(t) := [\nabla_x^2 S(\phi_z(t), t)]^{-1}$ along the instanton relates neighboring solutions of Hamilton's equations that start at $x_0$ through $\delta x(t) = Q(t) \, \delta p(t)$, and satisfies the forward Riccati equation
\begin{align}
\dot{Q} &= \nabla^2 \Psi(p_z) + \nabla b(\phi_z) Q + Q \nabla b(\phi_z)^\top \nonumber\\
&\quad + Q \, \nabla_x^2 \langle p_z, b(\phi_z) \rangle \, Q\,, \qquad Q(0) = 0\,.
\label{eq:riccati}
\end{align}
For linear drift, the last term vanishes, and Eq.~\eqref{eq:riccati} is the Lyapunov equation for the covariance of the fluctuations $(X^\eps - \phi_z) / \sqrt{\eps}$ under the tilted measure $\mathbb{Q}$. Compared to the Gaussian case~\cite{schorlepp-grafke-grauer:2023,grafke-schaefer-vanden-eijnden:2023}, the only change is $\nabla^2 \Psi(p_z)$ in place of the constant noise covariance (for additive noise with $\sigma \neq \Id$, it is replaced by $\sigma \nabla^2 \Psi(\sigma^\top p_z) \sigma^\top$). For a linear observable $f(x) = \langle g, x \rangle$, a Laplace approximation of the resulting sharp asymptotics of the PDF of $X^\eps_T$~\cite{schorlepp-grafke-grauer:2021,grafke-schaefer-vanden-eijnden:2023} gives the PDF prefactor
\begin{align}
C(z) = \frac{ \exp\left\{\frac{1}{2} \int_0^T \trace\left[ \nabla_x^2 \langle p_z, b(\phi_z) \rangle \, Q \right] \dd t\right\}}{\sqrt{\langle g, Q(T) g \rangle}}
\label{eq:riccati-prefactor}
\end{align}
with $\langle g,Q(T)g\rangle^{-1}=I''(z)$. Nonlinear and vector-valued observables are treated in Appendix~\ref{app:multi-constraint}. If $Q$ develops a pole, the prefactor can instead be evaluated from the fundamental matrix of the linearized Hamiltonian flow~\cite{schorlepp-grafke-grauer:2023}.

\subsection{Analytical example: one-dimensional gradient system at long times}
\label{sec:1d-analytic}

To gain intuition on how non-Gaussian noise affects the prefactor, consider the one-dimensional gradient system $\dd X^\eps_t = -V'(X^\eps_t) \dd t + \dd Z^\eps_t$, where $V$ is smooth with non-degenerate minimum at $x_a$. We are interested in the PDF $\rho_\eps(x_b)$ of $X^\eps_T$ at a point $x_b \neq x_a$ in its basin of attraction in the limit $T\to\infty$. Assume that the instanton is continuous and nonsaturated. In the long-time limit, the Hamiltonian $H(x,p) = -V'(x) p + \Psi(p)$ vanishes along the instanton, $E_z = 0$. Hence, $V'(\phi) = \Psi(p)/p$, which yields $\dot{\phi} = \Psi'(p) - V'(\phi) = \Psi'(p) - \Psi(p) / p$ and $\dot{p} = V''(\phi) p$. 
Since the time-translation mode $(\delta x,\delta p)=(\dot\phi,\dot p)$ solves the linearized Hamiltonian flow, the ratio $Q=\dot\phi/\dot p$ solves the Riccati equation~\eqref{eq:riccati},
\begin{align}
Q(t) = \frac{\dot{\phi}(t)}{\dot{p}(t)} = \frac{p \Psi'(p) - \Psi(p)}{p^2 \, V''(\phi)} = \frac{L(\eta_z(t))}{p_z(t)^2 \, V''(\phi_z(t))}\,,
\label{eq:1d-riccati-solution}
\end{align}
where we used $L(\eta_z) = p_z \eta_z - \Psi(p_z)$ with $\eta_z = \Psi'(p_z)$. It tends to the stationary variance $\Psi''(0) / (2 V''(x_a))$ of the linearized dynamics as $\phi_z \to x_a$, which is the appropriate initial condition for $T \to \infty$. Furthermore, the integrand in Eq.~\eqref{eq:riccati-prefactor} becomes $\frac12 \trace[\nabla_x^2 \langle p, b \rangle Q] = -\frac12 V'''(\phi) p Q = -\frac12 \frac{\dd}{\dd t} \log V''(\phi)$, since $p Q = \dot{\phi} / V''(\phi)$, such that the exponential factor in Eq.~\eqref{eq:riccati-prefactor} is $[V''(x_a) / V''(x_b)]^{1/2}$. With $\lambda = p_z(T)$ and $Q(T)^{-1/2} = |\lambda| V''(x_b)^{1/2} / L(\eta_z(T))^{1/2}$, we obtain
\begin{align}
\rho_\eps(x_b) &\overset{\eps \downarrow 0}{\sim} \sqrt{\frac{V''(x_a)}{2 \pi \eps}} \frac{|\lambda|}{\sqrt{L(\eta_z(T))}} \exp\left\{-\frac{I(x_b)}{\eps}\right\} \nonumber\\
&= \sqrt{\frac{V''(x_a)}{2 \pi \eps} \frac{|I'(x_b)|}{|\dot{\phi}_z(T)|}} \exp\left\{-\frac{I(x_b)}{\eps}\right\}\,,
\label{eq:1d-prefactor}
\end{align}
where the second form uses $L(\eta_z) = p_z (\eta_z - V'(\phi_z)) = p_z \dot{\phi}_z$ by  $H = 0$. For Gaussian noise $\Psi(p) = D p^2 / 2$, we have $I(x) = 2 (V(x) - V(x_a)) / D$ and $\dot{\phi}_z = V'(\phi_z)$, such that the prefactor $\sqrt{2 V''(x_a) / D}$ is independent of $x_b$, as it has to be for the Boltzmann distribution, cf.\ Ref.~\cite{schorlepp-grafke-grauer:2021}. For non-Gaussian noise, this cancellation does not occur in general, and the prefactor depends on the endpoint. Where $V''(x_b)\neq0$, it can also be written as $C(x_b)=\sqrt{V''(x_a)I''(x_b)/V''(x_b)}$. If $x_b$ approaches the boundary of the basin of attraction, i.e., a local maximum $x_s$ of~$V$, then $p_z(T) \to 0$ and $|\lambda| / \sqrt{L(\eta_z(T))} \to \sqrt{2 / \Psi''(0)}$, which is the Gaussian value for the same variance. This is consistent with the previously reported Gaussian escape-rate prefactor for continuous instantons~\cite{baule-sollich:2023}.

\subsection{Beyond continuous, non-degenerate instantons}
\label{sec:jump-prefactor}

The prefactor expressions of Sec.~\ref{sec:fredholm} rely on the same structure as in the Gaussian case: a continuous instanton with $p_z$ in the interior of $\dom \Psi$, and a strictly positive constrained second variation, such that fluctuations are Gaussian to leading order. For noise with bounded $\dom \Psi$, this structure can break down in several ways, and a complete theory is beyond the scope of this paper. Here, we only give partial results and comments for the two situations encountered in our examples, namely jumping instantons and saturated arcs for type C noise.

\paragraph*{Jumping instantons.}
If the instanton jumps at time $t_{\rm j}$, the momentum reaches the boundary of $\dom \Psi$ there, and $\nabla^2 \Psi(p_z(t))$ diverges as $t \to t_{\rm j}$. For type C noise, $\psi_\alpha''(u) \sim \frac12 (1 - u)^{\alpha - 2}$ as $u \uparrow 1$, such that $\nabla^2 \Psi(p_z(t)) \asymp |t - t_{\rm j}|^{\alpha - 2}$ assuming the boundary is approached linearly in time. This singularity is integrable for $\alpha \in (1,2)$, but not for $\alpha \in (0,1)$, and in either case, the Fredholm and Riccati expressions above are not directly applicable. A natural heuristic idea is to split the path at the jump: for a single jump of size $\Delta q=q_+-q_->0$ at time $t_{\rm j}$ in a scalar coordinate, we write
\begin{align}
\rho_\eps(z) &\approx \int_0^T \dd t_{\rm j} \int \dd q_- \int \dd q_+ \; \rho^{(1)}_\eps(q_-, t_{\rm j}) \nonumber\\
&\quad \times \nu_\eps(q_+ - q_-) \, \rho^{(2)}_\eps(z; q_+, t_{\rm j})\,,
\label{eq:jump-splitting}
\end{align}
where $\rho^{(1)}_\eps(q_-, t_{\rm j})$ is the transition density to reach $q_-$ at time $t_{\rm j}$, $\rho^{(2)}_\eps(z; q_+, t_{\rm j})$ is the PDF of $f(X^\eps_T)$ when starting from $q_+$ at time $t_{\rm j}$, and $\nu_\eps$ is the density of the L{\'e}vy measure of $Z^\eps$, which accounts for a single large noise jump in between. For the type C example, $\nu_\eps(\Delta q) = \eps^{\alpha - 1} K_\alpha(\Delta q) \, e^{-\Delta q / (a \eps)}$ with $K_\alpha(\Delta q) = r a^\alpha \Delta q^{-1-\alpha} / (2 \Gamma(2 - \alpha))$. Before and after the jump, the densities are described by continuous instantons, $\rho^{(i)}_\eps \sim (2 \pi \eps)^{-1/2} C_i e^{-S_i/\eps}$ for $i = 1,2$, with actions $S_1, S_2$ and prefactors $C_1, C_2$ that can be computed as in Sec.~\ref{sec:fredholm}. The exponential factor of $\nu_\eps$ contributes the jump cost $L_\infty(\Delta q) = \Delta q / a$. Minimizing the total action $S = S_1(q_-, t_{\rm j}) + (q_+ - q_-)/a + S_2(q_+, t_{\rm j})$ over the matching point $(t_{\rm j}, q_-, q_+)$ reproduces the corner conditions of Sec.~\ref{sec:jumps}: the momentum is continuous across the jump and takes the boundary value $1/a$, and the Hamiltonian is conserved. Evaluating Eq.~\eqref{eq:jump-splitting} by Laplace's method around an isolated, non-degenerate matching point $(t_{\rm j}^*, q_-^*, q_+^*)$ then formally gives
\begin{align}
\rho_\eps(z) &\overset{\eps \downarrow 0}{\sim} \sqrt{2 \pi} \, \eps^{\alpha - 1/2} \frac{C_1 \, K_\alpha(\Delta q^*) \, C_2}{\sqrt{\det \nabla^2 S}} \exp\left\{-\tfrac{I(z)}{\eps}\right\}\;\;\;
\label{eq:jump-prefactor}
\end{align}
with $\Delta q^* = q_+^* - q_-^*$ and the $3 \times 3$ Hessian $\nabla^2 S$ of the total action at the matching point, which can be assembled from Riccati solutions on both segments and the instanton velocities at the jump. We give the details, as well as the generalization to $n$ dimensions, in Appendix~\ref{app:jump-prefactor}. Compared to continuous instantons~\eqref{eq:sharp-asymptotics}, the prefactor carries an anomalous relative factor $\eps^\alpha$. This agrees with the prefactor of the escape rate found in Ref.~\cite{baule-sollich:2023} in one dimension, and reflects the probability of a single large noise jump of size $\Delta q^*$.

The splitting~\eqref{eq:jump-splitting} is heuristic and has limitations. In the two-dimensional example of Sec.~\ref{sec:2d}, we find that Eq.~\eqref{eq:jump-prefactor} approaches the direct sampling results for $\alpha = 3/2$ sufficiently far from the onset of the jump, but not for $\alpha = 1/2$. There, the nonintegrable singularity of $\psi_\alpha''$ at the jump implies that the curvature of $S_1$ vanishes and the instanton velocity diverges at the matching point, such that the Gaussian approximation of the segments around it is inconsistent. Furthermore, as the jump size $\Delta q^* \to 0$ at the onset of the jump, $K_\alpha(\Delta q^*)$ diverges, and Eq.~\eqref{eq:jump-prefactor} is not uniform in $z$. A uniform prefactor through the onset of jumps would require a matched asymptotic boundary layer analysis, which we leave for future work.

\paragraph*{Saturated arcs.}
Another degeneracy occurs if the instanton is continuous, but contains a saturated arc for type C noise with $\alpha \in (1,2)$, on which $\nabla^2 L(\eta_z) = 0$. The fluctuations around the instanton are governed by the exponentially tilted L{\'e}vy measure, $e^{\langle p_z, y \rangle} \nu(\dd y)$ in unscaled variables, cf.\ Eq.~\eqref{eq:tilt}, which, for $p_z$ on the boundary of $\dom \Psi$, has power-law tails $\propto |y|^{-1-\alpha}$ with infinite variance. Fluctuations on a saturated arc are hence not Gaussian, and the power of $\eps$ in the prefactor changes. Using the finite-dimensional caricature $P_\eps(z) := (2\eps)^{-N} \int_{\RR^N} \mathds{1}_{\{F(\eta) \geq z\}} e^{-\norm{\eta}_1 / \eps} \dd^N \eta$, one can show that a Laplace expansion around a minimizer with $k$ nonzero components yields $P_\eps(z) \propto \eps^{-(k-1)/2} e^{-I(z)/\eps}$. In this case, the cost is flat along the $k-1$ directions, and fluctuations along them are confined only by the curvature of the target set. In the continuum, the number of such components grows with the temporal resolution of the saturated arc, so that a prefactor of the form~\eqref{eq:sharp-asymptotics} cannot be expected. In the examples below, we do not assign prefactors to instantons with saturated arcs.

%%%%%%%%%%%%%%%%%%%%%%%%%%%%%%%%%%%%%%%%%%%%%%%%%%%%%%%%%%%%%%%%%%%%%%%%%%%
\section{Examples}
\label{sec:examples}
%%%%%%%%%%%%%%%%%%%%%%%%%%%%%%%%%%%%%%%%%%%%%%%%%%%%%%%%%%%%%%%%%%%%%%%%%%%

\begin{figure*}
\centering
\includegraphics[width=\textwidth]{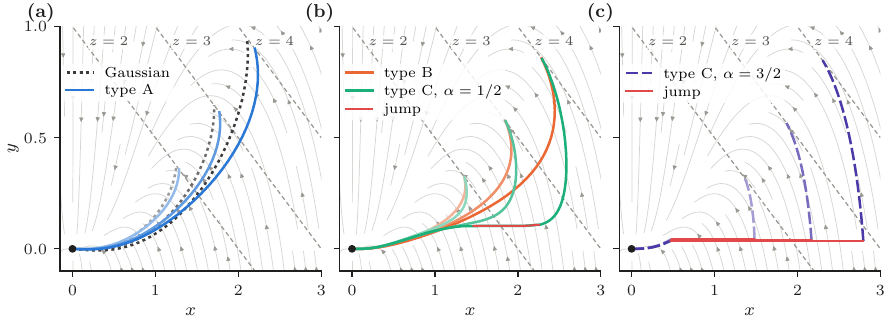}
\caption{Instantons of the two-dimensional model~\eqref{eq:2d-model} for $z \in \{2, 3, 4\}$, with the target sets $x + 2y = z$ (dashed). The vector field corresponds to the deterministic drift. (a)~Gaussian and type A instantons from fixed-time control optimization~\eqref{eq:control-problem}, (b)~TR solutions for type B and type C ($\alpha = 1/2$) instantons from~\eqref{eq:tr-problem}, (c)~type C ($\alpha = 3/2$) TR plus shooting-refined instantons. Jumps are shown in red: for $\alpha = 1/2$, the instanton for $z = 4 > z_{\rm c}$ jumps, while for $\alpha = 3/2$ all shown instantons in (c) jump.}
\label{fig:2d-instantons}
\end{figure*}

We now apply the methods of Secs.~\ref{sec:control} and~\ref{sec:prefactor} to three numerical examples. In Sec.~\ref{sec:2d}, we consider a two-dimensional nonlinear SDE, for which we compare all noise types, compute jumping instantons, and test the prefactor estimates against direct sampling. In Sec.~\ref{sec:rtp}, we turn to an RTP in a harmonic potential, in which the L{\'e}vy noise acts only on the orientation. Finally, in Sec.~\ref{sec:gsebm}, we consider a spatially extended climate model with multiplicative noise, and compute transitions between its metastable (warm and snowball) states.

\subsection{Two-dimensional nonlinear SDE}
\label{sec:2d}

As a first example, we consider a two-dimensional SDE~\eqref{eq:sde} with additive noise used in Refs.~\cite{schorlepp-grafke-grauer:2021,schorlepp-tong-grafke-etal:2023},
\begin{align}
b(x,y) &= \begin{pmatrix} -x - x y\\ -4y + x^2 \end{pmatrix}\,, \quad x_0 = (0,0)\,, \nonumber\\
T &= 1\,, \quad f(x,y) = x + 2y\,,
\label{eq:2d-model}
\end{align}
for which we compare four noise types with the same covariance $\nabla^2 \Psi(0) = \operatorname{diag}(1, 1/4)$: Gaussian, $\Psi(p) = \tfrac12 p_1^2 + \tfrac18 p_2^2$, and types A, B and C with $\Psi(p) = \psi(p_1) + \psi(p_2/2)$, i.e., $r = (1,1)$ and $a = (1, 1/2)$ in Eq.~\eqref{eq:scalar-noise} (amplitude convention), such that all differences between the noise types are due to higher cumulants. For comparison, we also compute type A and B rate functions with the same covariance obtained by scaling the jump intensity instead of the amplitude, $\Psi(p) = \psi(p_1) + \frac14 \psi(p_2)$. For numerical details, see Appendix~\ref{app:2d-num}.

\begin{figure*}
\centering
\includegraphics[width=0.85\textwidth]{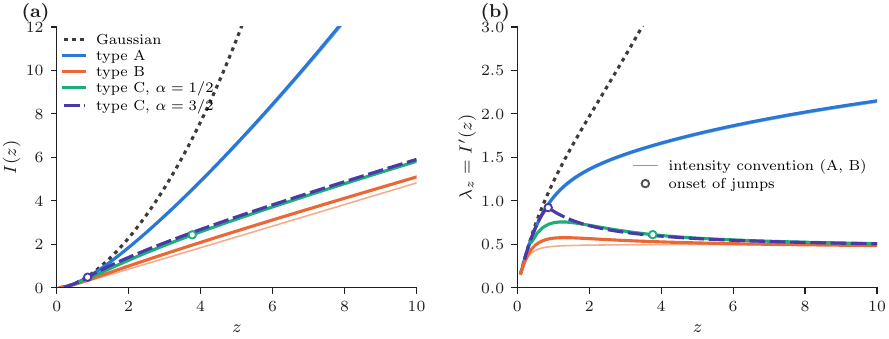}
\caption{(a)~Rate functions $I(z)$ and (b)~their derivatives $\lambda_z = I'(z)$ for Eq.~\eqref{eq:2d-model} for all noise types, in the amplitude convention $\Psi(p) = \psi(p_1) + \psi(p_2/2)$. For small~$z$, all rate functions coincide, since the noise types share the same covariance. For larger~$z$, the rate function under Gaussian noise grows approximately quadratically, the type A rate function superlinearly but more slowly, and those of types B and C asymptotically linearly, with $\lambda_z$ approaching a common value. For types B and C, $\lambda_z$ first increases, reaches a maximum and then decreases towards this value, i.e., $I$ is concave for large $z$, and for type C with $\alpha = 3/2$, $\lambda_z$ has a corner at the onset of jumps. Open circles mark this onset for type C, at $z_{\rm c} \approx 3.77$ ($\alpha = 1/2$) and $z_{\rm c} \approx 0.86$ ($\alpha = 3/2$). Thin light lines show the intensity convention $\psi(p_1) + \frac14 \psi(p_2)$ for types A and B (indistinguishable for type A at this scale).}
\label{fig:2d-rates}
\end{figure*}

\begin{figure*}
\centering
\includegraphics[width=\textwidth]{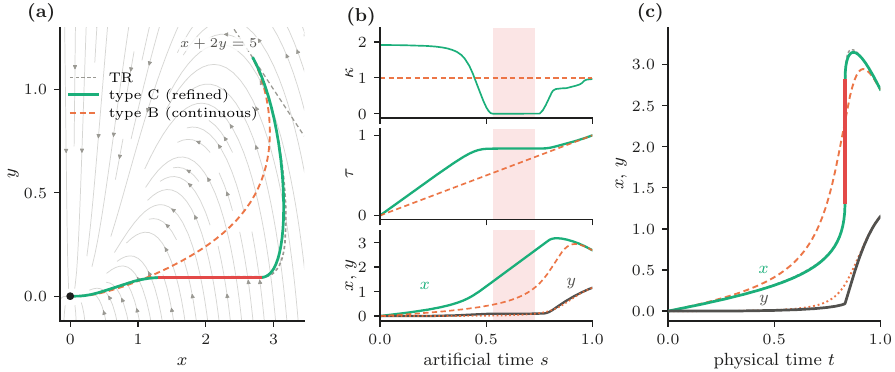}
\caption{Type C ($\alpha = 1/2$) instanton of the two-dimensional model at $z=5$, obtained by TR control optimization~\eqref{eq:tr-problem} and then refined by shooting (green), compared with the continuous type B solution (orange dashed). (a)~Instanton in the $(x,y)$-plane; thin gray dashed: the TR solution before refinement. (b)~Time density $\kappa(s)$, physical time $\tau(s)$, and state components of the TR solution. (c)~State components in physical time; for type B, $x$ is dashed and $y$ dotted.}
\label{fig:2d-jump}
\end{figure*}

\paragraph*{Instantons.}
Gaussian and type A instantons, obtained by fixed-time control optimization~\eqref{eq:control-problem}, are continuous and qualitatively similar, see Fig.~\ref{fig:2d-instantons}(a). The type A noise, however, makes large forcing cheaper, which lowers the action considerably compared to Gaussian noise. For types B and C, the TR solutions of Eq.~\eqref{eq:tr-problem} in Fig.~\ref{fig:2d-instantons}(b,c) have even lower actions, and the rate functions grow asymptotically linearly in $z$, see Fig.~\ref{fig:2d-rates}. While type B instantons remain continuous for all $z$, as predicted in Sec.~\ref{sec:jumps}, the type C instanton first moves predominantly in the horizontal direction, where the recession cost $L_\infty(\eta) = |\eta_1| + 2|\eta_2|$ is cheaper, to a region where the $x^2$ term in the drift assists the subsequent vertical motion. For $\alpha = 1/2$ and $z > z_{\rm c} \approx 3.77$, this horizontal excursion develops into a genuine jump, located by the TR formulation~\eqref{eq:tr-problem} as an interval of artificial time with $\kappa = 0$ and finite displacement, see Fig.~\ref{fig:2d-jump}. For $\alpha = 3/2$, the instantons jump already for $z > z_{\rm c} \approx 0.86$.

For $\alpha = 1/2$, we find that the jump size grows as $(z - z_{\rm c})^{1/2}$, and $I$ is continuously differentiable across the onset, without a detectable kink in $\lambda_z$. For $\alpha = 3/2$, the jump size grows linearly in $z - z_{\rm c}$, and $\lambda_z$ is continuous but has a corner there. In all cases, $H$ is constant on the continuous segments, and $H(q, p_z) \leq E_z$ along the jump, as required by condition~(ii) of Sec.~\ref{sec:jumps}.

\begin{figure*}
\centering
\includegraphics[width=\textwidth]{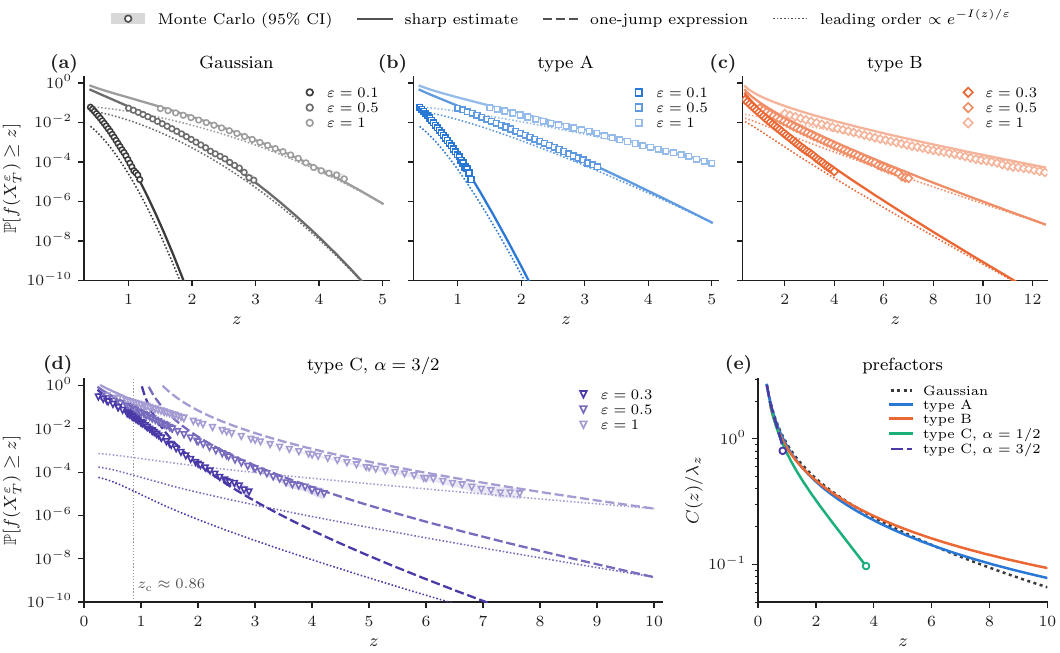}
\caption{Tail probabilities $\PP[f(X^\eps_T) \geq z]$ of the two-dimensional SDE~\eqref{eq:2d-model} for (a)~Gaussian, (b)~type A, (c)~type B and (d)~type C ($\alpha = 3/2$) noise. Markers with shaded $95\%$ Wilson confidence intervals show direct Monte Carlo estimates from explicit Euler simulations ($10^6$ paths per $\eps$, points with fewer than $10$ hits omitted). Solid lines: sharp estimates~\eqref{eq:sharp-asymptotics}; dotted lines: leading order $\text{const.} \times \exp\{-I(z)/\eps\}$, normalized to the sharp estimate at the largest $z$ shown. In (d), dashed lines are the one-jump expression~\eqref{eq:jump-prefactor} for $z > z_{\rm c} \approx 0.86$. Where the instanton contains a well-developed jump, it approaches the simulations as $\eps$ decreases. Close to the onset, the jump of the instanton is small, and the one-jump expression overestimates the simulations, since it diverges as the jump size tends to zero. (e)~Prefactors $C(z)/\lambda_z$ of continuous instantons for all noise types; open circles mark the last plotted continuous-branch values below the jump onset for type C.}
\label{fig:2d-prob}
\end{figure*}

\paragraph*{Prefactors and comparison to Monte Carlo.}
We compute prefactors~\eqref{eq:sharp-asymptotics} and~\eqref{eq:jump-prefactor} in the Riccati formulation in physical time~\eqref{eq:riccati-prefactor} (and cross-check them with the Fredholm formulation~\eqref{eq:fredholm-prefactor} for Gaussian and type A noise). Figure~\ref{fig:2d-prob} compares the resulting tail probability estimates to direct Monte Carlo simulations of Eq.~\eqref{eq:sde} at different $\eps$. For these simulations, type A and B noise are simulated exactly as compound Poisson processes. For the infinite-activity type C noise, jumps $\eps a_i y$ of $Z^\eps$ with $|y| \geq 0.1$ in units of the jump density $\omega$ are sampled exactly by rejection from a Pareto proposal, while the remaining small jumps are replaced by a Gaussian increment of the same variance~\cite{asmussen-rosinski:2001}.

For Gaussian, type A and type B noise, the sharp estimates agree with sampling as $\eps$ decreases, and including the prefactor consistently improves upon the leading-order estimate $\text{const.} \times \exp\{-I(z)/\eps\}$. The prefactor varies by orders of magnitude over the range of $z$ considered, and cannot be absorbed into a constant, cf.~Refs.~\cite{schorlepp-tong-grafke-etal:2023,schorlepp-grafke:2025}. Interestingly, the prefactors for Gaussian and type A noise are close in this example, despite the very different actions, see Fig.~\ref{fig:2d-prob}(e). For type C with $\alpha = 3/2$, the instanton jumps for $z > z_{\rm c}$. There, the one-jump expression~\eqref{eq:jump-prefactor} approaches the Monte Carlo data as $\eps$ decreases away from the onset of the jump, but not uniformly in $z$, as expected for a dynamical phase transition. For $\alpha = 1/2$ (not shown), the continuous formula agrees with sampling below the onset of the jump, while naively applying the one-jump expression~\eqref{eq:jump-prefactor} overestimates the Monte Carlo data above the onset by factors that grow as $\eps$ decreases, confirming its failure for $\alpha \in (0,1)$ discussed in Sec.~\ref{sec:jump-prefactor}.

\subsection{Run-and-tumble particle in an anisotropic trap}
\label{sec:rtp}

As a second example, we consider an RTP in an anisotropic harmonic trap. Unlike active Brownian particles, whose orientation undergoes rotational diffusion, RTPs reorient through discrete tumbles~\cite{bechinger-dileonardo-lowen-etal:2016}. Although the two models can have similar typical, coarse-grained dynamics at matched persistence times~\cite{tailleur-cates:2008,cates-tailleur:2013,cates-tailleur:2015,solon-cates-tailleur:2015}, their behavior under confinement~\cite{khatami-etal:2016} or in rare events~\cite{cates-tailleur:2013,woillez-zhao-kafri-etal:2019,gradenigo-majumdar:2019,mori-ledoussal-majumdar-etal:2021} may differ significantly. In our scaling, typical orientational fluctuations become diffusive, while large deviations retain the tumble statistics.

\begin{figure*}
\centering
\includegraphics[width=\textwidth]{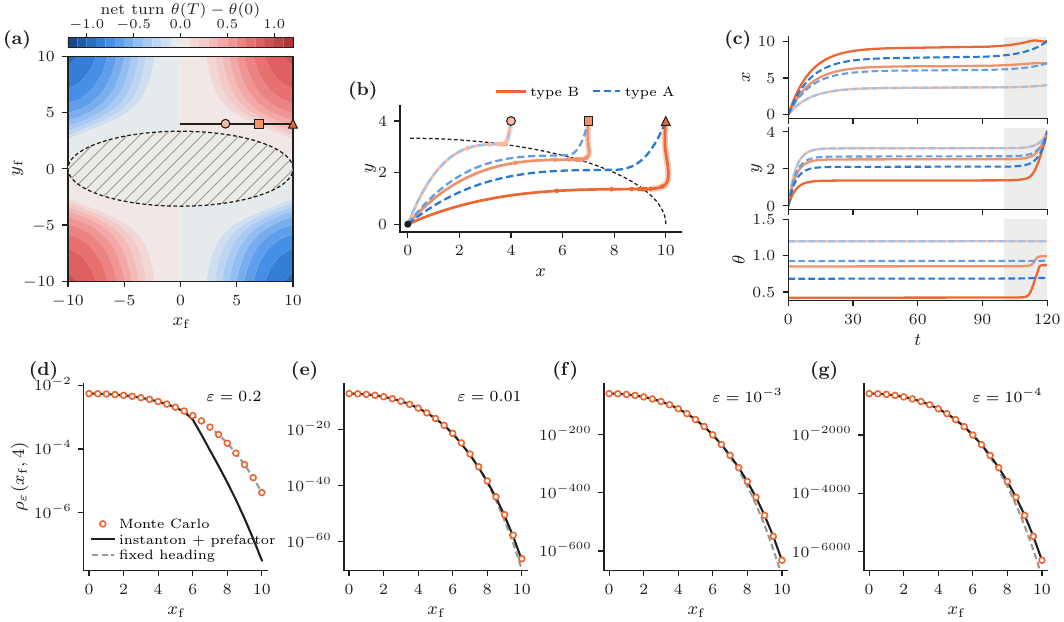}
\caption{RTP with type B tumbles (with type A for comparison in (b,c)). (a)~Net instanton turn as a function of endpoint; hatching marks the region inside the ellipse excluded from the present analysis. Markers indicate endpoints $(4,4)$, $(7,4)$ and $(10,4)$. (b)~The instantons' spatial paths, for type B (solid) and type A (dashed) tumbles. (c)~Positions and headings for both types, with shaded gray area indicating the last $20$ units of time where  the transition towards the endpoint happens. (d--g)~Final-position PDF along the slice $y_{\rm f}=4$ (black line in (a)) at the indicated $\eps$: importance-sampling estimates, sharp asymptotics~\eqref{eq:rtp-uniform} (instanton + prefactor), and the fixed-heading reference. At $\eps=0.2$, sampling follows the fixed-heading reference; as $\eps$ decreases it approaches the sharp estimate, including endpoints reached by turning instantons.}
\label{fig:rtp-typeB}
\end{figure*}

Specifically, we consider a particle with position $(x,y)$ and orientation $\theta$ in the trap,
\begin{align}
\begin{cases}
\dd X = \left(-k_x X + v_0 \cos \Theta \right) \dd t + \sqrt{\eps D} \, \dd W_x\,,\\
\dd Y = \left(-k_y Y + v_0 \sin \Theta \right) \dd t + \sqrt{\eps D} \, \dd W_y\,,\\
\dd \Theta = \dd Z^\eps_\theta\,,
\end{cases}
\label{eq:rtp}
\end{align}
where $\Psi_\theta(p)=r\psi(ap)$ describes the angular tumbles, with intensity scale $r$ and angular jump-amplitude scale $a$. We use $k_x=0.1$, $k_y=0.3$, $v_0=D=1$, $a=2$, $r=10^{-4}$ and $T=120$, following a suggestion by K.~Cheruvary, R.~D{\'i}az Hern{\'a}ndez Rojas and P.~Sollich~\cite{sollich-diazhernandezrojas:2026}. Starting at the origin $X_0 = Y_0 = 0$, we compute the final-position PDF $\rho_\eps(x_{\rm f},y_{\rm f})$ of $(X_T, Y_T)$ for type B, as well as instantons for type C ($\alpha=3/2$) tumbles, with type A tumbles as a reference.

\paragraph*{Reduced problem and prefactor.}
The initial and final headings $\Theta_0,\Theta_T$ are free variables in the instanton problem. We hence optimize over the initial heading and impose no constraint on the final heading. For a given orientation history $(\theta(t))_{t \in [0,T]}$, the spatial dynamics in Eq.~\eqref{eq:rtp} is linear with Gaussian noise, such that the spatial controls can be eliminated analytically. This yields the reduced optimization problem
\begin{align}
&I(x_{\rm f}, y_{\rm f}) = \min_{\theta} \left\{ \sum_{i \in \{x,y\}} \frac{(i_{\rm f} - \mu_i[\theta])^2}{2 \Sigma_i} + \int_0^T L_\theta(\dot{\theta}) \, \dd t \right\}\,, \nonumber\\
&\mu_i[\theta] =\textstyle \int_0^T e^{-k_i (T - t)} v_i(\theta(t)) \, \dd t\,, \;\; \Sigma_i = D \frac{1 - e^{-2 k_i T}}{2 k_i}\,,
\label{eq:rtp-reduced}
\end{align}
with $v_x(\theta)=v_0\cos\theta$, $v_y(\theta)=v_0\sin\theta$, and angular cost $L_\theta(\eta)=r\psi^*(\eta/(ra))$. The means $\mu_i[\theta]$ are the endpoints without spatial noise; the first, quadratic term in the objective in Eq.~\eqref{eq:rtp-reduced}, denoted $R[\theta]$ below, is the spatial-noise cost of the remaining displacement. The optimal spatial forcing is proportional to $e^{-k_i(T-t)}$ and acts mainly during the last relaxation times.

This reduction is exact also at the level of the prefactor. Conditional on $\theta$, $(X_T, Y_T)$ is Gaussian with density $(2 \pi \eps)^{-1} (\Sigma_x \Sigma_y)^{-1/2} \exp\{-R[\theta]/\eps\}$, and the remaining Laplace integral over the angular noise $\eta_\theta = \dot{\theta}$ is unconstrained. With $M_z$ as in Eq.~\eqref{eq:whitening} for the angular component, this yields the PDF conditional on the optimal initial angle $\theta_0^*$,
\begin{gather}
\rho_\eps(x_{\rm f}, y_{\rm f} \mid \theta_0^*) \overset{\eps \downarrow 0}{\sim} \frac{C_{\theta_0^*}}{2 \pi \eps} \exp\left\{-\frac{I(x_{\rm f}, y_{\rm f})}{\eps}\right\}\,, \nonumber\\
C_{\theta_0^*} = \frac{1}{\sqrt{\Sigma_x \Sigma_y}} \det\left(\Id + M_z \frac{\delta^2 R}{\delta \eta_\theta^2} M_z\right)^{-1/2}\,,
\label{eq:rtp-prefactor}
\end{gather}
in agreement with Eq.~\eqref{eq:multi-fredholm} for the full problem.
To obtain the PDF for a uniformly distributed initial heading, we average the conditional density over that heading. A further Laplace approximation to this integral yields
\begin{align}
\rho_\eps(x_{\rm f},y_{\rm f})\overset{\eps\downarrow0}{\sim}\frac{C(x_{\rm f},y_{\rm f})}{\sqrt{2\pi\eps}}e^{-I(x_{\rm f},y_{\rm f})/\eps}\,,
\label{eq:rtp-uniform}
\end{align}
where $
C= C_{\theta_0^*} / (2\pi\sqrt{\partial^2_{\theta_0}I_{\theta_0}|_{\theta_0=\theta_0^*}})$. Here, $I_{\theta_0}$ denotes the reduced action minimized at fixed initial heading and fixed spatial endpoint, and $I=\min_{\theta_0}I_{\theta_0}$. The curvature is evaluated at the optimal initial heading $\theta_0^*$. The initial-angle integral gives a factor $\sqrt{2\pi\eps}$, explaining the $\eps^{-1/2}$ scaling of the two-dimensional PDF. All numerical details are given in Appendix~\ref{app:rtp-numerics}. We minimize the reduced action in physical time, since we found no indication of jumps during searches with TR.

\paragraph*{Instantons.}
To understand the instantons' qualitative behavior in Fig.~\ref{fig:rtp-typeB}(b,c), we note that the deterministic equilibria $(v_0\cos\theta_0/k_x,v_0\sin\theta_0/k_y)$ form an ellipse where the long-time rate function vanishes (black dashed line in Fig.~\ref{fig:rtp-typeB}(a,b)). The unequal relaxation times, $1/k_x=10$ and $1/k_y\approx3.3$, give the final coordinates different temporal weightings of the heading. A turn at constant rate through $\Delta\theta$ over time $\tau$ costs $\tau r\psi^*(\Delta\theta/(\tau r a))$. At low intensity ($r = 10^{-4}$ here), even small turns probe the large-control regime, where types B and C are cheaper than Gaussian noise of equal variance. The instanton balances this turning cost against spatial fluctuations. Type A's superlinear turning cost favors an almost constant heading and greater reliance on spatial noise.

Type B instantons use two strategies, see Fig.~\ref{fig:rtp-typeB}(a--c). After relaxing to the equilibrium set by their initial heading, they realize the rare event during the final relaxation times. Near the ellipse or the $y$-axis, e.g., at $(x_{\rm f}, y_{\rm f}) = (4,4)$, the heading stays constant, $\theta(t) \equiv \theta_0$, and spatial noise supplies the displacement. Farther out, as at $(7,4)$ and $(10,4)$, the particle keeps a shallow heading and turns toward the $y$-direction late. Because $X$ relaxes more slowly, it retains displacement accumulated before the turn, while $Y$ responds more quickly to the new heading, producing an almost vertical final run. The turn grows continuously with distance from the ellipse and remains continuous in time, as predicted in Sec.~\ref{sec:jumps}. Our focus here is on rare excursions beyond the ellipse. We therefore exclude the hatched region in its interior from the present analysis, leaving possible competition between candidate instantons there for future investigation. Type A instantons coincide with type B at $(4,4)$, but barely turn at $(7,4)$ and $(10,4)$ and rely on spatial noise instead, see Fig.~\ref{fig:rtp-typeB}(b,c).

For type C with $\alpha=3/2$ we found that the same spatial strategies occur (not shown). In contrast to type B, the angular cost becomes affine at large enough $|\dot\theta|$, thus allowing for -- and indeed displaying -- saturated arcs. We leave the study of these arcs (and their prefactors) for future work, and refer to the forthcoming Ref.~\cite{sollich-diazhernandezrojas-paper:2026} for a detailed analysis in the present RTP problem.

\paragraph*{Monte Carlo simulations.}
To sample $\rho_\varepsilon(x_{\rm f}, y_{\rm f})$, we take the initial heading $\Theta_0 \sim {\cal U}([0, 2 \pi))$ and leave the final heading unrestricted, consistent with the initial-angle average in Eq.~\eqref{eq:rtp-uniform}. We also use the reduced formulation for Monte Carlo: we sample angular histories, evaluate the conditional Gaussian position density analytically, and integrate numerically over the uniform initial heading. For type B tumbles, we use importance sampling with the instanton tilt~\eqref{eq:tilt} to resolve rare turning histories. As a reference, we also consider the ``fixed-heading ensemble'', where  $\Theta_t \equiv\Theta_0$, and the displacement is the uniform mixture of conditional Gaussian densities.

As Fig.~\ref{fig:rtp-typeB}(d--g) shows, along the slice $y_{\rm f}=4$, sampling agrees well with the sharp estimate for $\eps\leq0.01$, while the fixed-heading reference increasingly underestimates the PDF beyond $x_{\rm f}\approx6$. At $\eps=10^{-4}$, sampling and the sharp estimate differ by at most a few percent. These endpoints thus indeed favor late-turning histories. At comparatively large noise strength $\eps=0.2$, the expected number of tumbles is only $rT/(2\eps)=0.03$. Sampling then instead follows the fixed-heading reference and exceeds the asymptotic estimate by two orders of magnitude at $(10,4)$.

\subsection{Ghil--Sellers energy balance model with L{\'e}vy-perturbed solar forcing}
\label{sec:gsebm}

As a spatially extended example, we consider the Ghil--Sellers EBM~\cite{sellers:1969,ghil:1976} with the parameterization and corrections of B{\'o}dai et al.~\cite{bodai-lucarini-lunkeit-etal:2015}, following the stochastic setting of Lucarini et al.~\cite{lucarini-serdukova-margazoglou:2022}. The zonally averaged surface temperature $\mathcal{T}(x,t)$, with normalized latitude $x\in[-1,1]$ and time in years, satisfies
\begin{align}
\mathcal{C}(x)\partial_t\mathcal{T}&=\mathcal{D}[\mathcal{T}]+(\mu+\dot Z^\eps_t)\mathcal{Q}(x)[1-\mathcal{A}(x,\mathcal{T})]-\mathcal{R}(\mathcal{T})\,,
\label{eq:gsebm}
\end{align}
with heat capacity $\mathcal C$, meridional heat transport $\mathcal D$, insolation $\mathcal Q$, temperature-dependent albedo $\mathcal A$, outgoing radiation $\mathcal R$, and relative solar irradiance $\mu=1.05$.
The scalar noise perturbs solar irradiance, giving the multiplicative coupling $\sigma(\mathcal T)=\mathcal Q(x)[1-\mathcal A(x,\mathcal T)]/\mathcal C(x)$ in the notation of Eq.~\eqref{eq:sde}. We use the Marcus interpretation, consistent with the TR formulation for jumps~\eqref{eq:tr-problem}, and discretize latitude into 201 finite-volume cells. There are two metastable climates, warm (W) and snowball (SB). Their basins of attraction are separated by the stable manifold of an unstable saddle equilibrium M, also called the edge state. For our noise laws, we use $r=0.01$, $a=2$, and variance $0.04\eps$ per year, matched to the Gaussian reference (and referred to as the ``canonical'' choice below). We compute and compare instantons for transitions W $\to$ SB and SB $\to$ W.
As in Ref.~\cite{lucarini-serdukova-margazoglou:2022}, we visualize paths using the area-weighted global mean $\langle\mathcal T\rangle$ and the equator-to-pole difference $\Delta\mathcal T$, defined as the area-weighted mean over $|x|<1/3$ minus that over $|x|>1/3$. In contrast to Ref.~\cite{lucarini-serdukova-margazoglou:2022}, we do not consider $\alpha$-stable noise, whose infinite CGF lies outside the present setting and whose escape rates scale polynomially~\cite{imkeller-pavlyukevich:2006}.

\paragraph*{Finite-window transition problem.}
Transitions in the stationary regime leave the source and approach the target over infinite time (i.e., for $T \to \infty$). We approximate this infinite-time problem by allowing control over a finite physical-time window of duration $T_{\rm c} = 20$ years, starting at the source state and requiring only that the endpoint cross the stable manifold of M. Subsequent relaxation to the target is deterministic and costs nothing, so the control need not act throughout the full transition. We solve the resulting TR problem~\eqref{eq:tr-problem} with SLSQP~\cite{kraft:1988} and JAX gradients, and verify that each optimized control produces a transition through an independent full-transition simulation. Restricting the control window overestimates the infinite-time escape action, but we checked that doubling the window lowered the W$\to$SB actions by only about $0.1\%$ for type B and type C with $\alpha=3/2$. Table~\ref{tab:gsebm} compares the resulting actions with those obtained using fixed-time annual controls over 120 years. The actions differ by less than $8\%$ across these calculations, although this comparison changes both the control-window duration and the temporal representation of the control.

\begin{table}
\caption{Ghil--Sellers transition actions for different noise types of equal variance. TR: time reparameterization with a 20-year control window. Fixed time: annual controls over 120 years, taking the lowest action across multiple initializations.}
\label{tab:gsebm}
\begin{ruledtabular}
\begin{tabular}{lcccc}
 & \multicolumn{2}{c}{$I_{\rm W\to SB}$} & \multicolumn{2}{c}{$I_{\rm SB\to W}$}\\
noise & TR & fixed time & TR & fixed time\\
\hline
Gaussian & 3.595 & 3.473 & 2.564 & 2.489\\
type A & 1.699 & 1.693 & 1.320 & 1.314\\
type B & 0.541 & 0.549 & 0.452 & 0.455\\
type C, $\alpha=1/2$ & 0.596 & 0.622 & 0.508 & 0.538\\
type C, $\alpha=3/2$ & 0.602 & 0.632 & 0.515 & 0.556
\end{tabular}
\end{ruledtabular}
\end{table}

\begin{figure*}
\centering
\includegraphics[width=\textwidth]{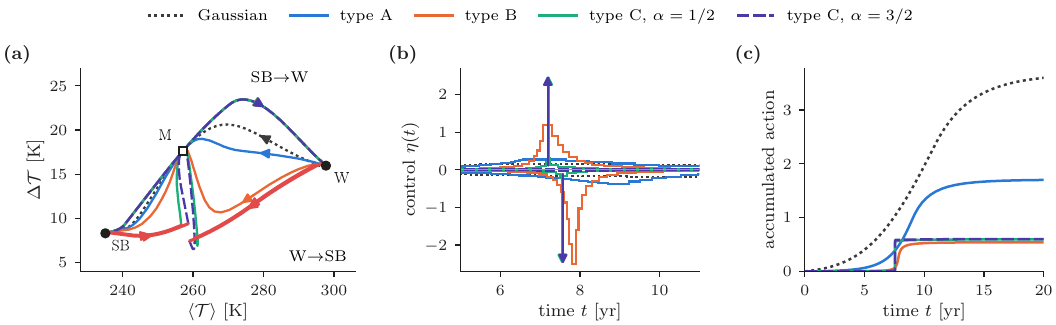}
\caption{Ghil--Sellers EBM. (a)~TR instantons between the warm (W) and snowball (SB) states for all noise types, projected to $\langle \mathcal{T} \rangle$ and $\Delta \mathcal{T}$. The square marks the edge state M, arrows indicate the direction of the transitions, and jumps of the type C instantons are shown in red. (b)~Controls $\eta(t)$ in part of the control window (negative for W$\to$SB, positive for SB$\to$W); the type C jumps are shown as arrows (off scale). (c)~Accumulated action of the W$\to$SB instantons, cf.\ Table~\ref{tab:gsebm}.}
\label{fig:gsebm}
\end{figure*}

\begin{figure*}
\centering
\includegraphics[width=\textwidth]{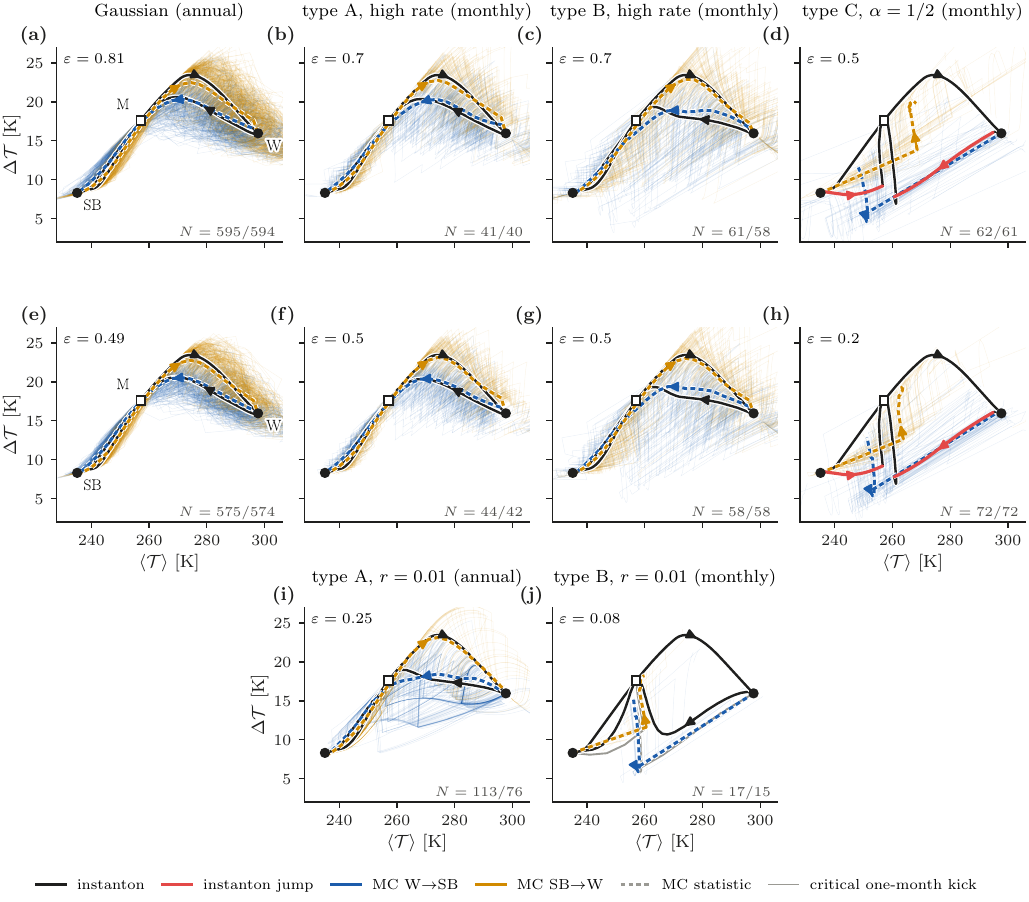}
\caption{Ghil--Sellers transitions, Monte Carlo transition paths (colored paths), and instantons (black) in the $(\langle\mathcal T\rangle,\Delta\mathcal T)$ plane for different noise types (columns). Dashed paths are conditional means from Monte Carlo, except for type C and canonical type B, which use medians aligned at the largest monthly increment. (a--h)~Two noise strengths for Gaussian with canonical forcing, high-rate types A/B ($r=0.64$, $a=0.25$), and canonical type C ($\alpha=1/2$), respectively. (i,j)~Canonical types A/B ($r=0.01$). In (j), the thin gray curves show the critical one-month-kick paths~\cite{lucarini-serdukova-margazoglou:2022} followed by canonical type B transitions rather than the instantons. In (d,h), type C transitions realize the instanton jump through a large monthly increment, with an overshoot. The stated $N$ gives observed W$\to$SB/SB$\to$W transition counts.}
\label{fig:gsebm-mc}
\end{figure*}

\paragraph*{Instantons.}
We find that all instanton paths approach the saddle state M before relaxing to the target climate, see Fig.~\ref{fig:gsebm}(a). The forcing that drives this approach differs across noise types: Gaussian and type A instantons use broad forcing, type B uses a short continuous pulse, and type C contains a jump, see Fig.~\ref{fig:gsebm}(b). During a Marcus jump, physical time is frozen: only the noise coupling $\sigma({\cal T})$ changes the temperature, while the heat-transport term $\mathcal D[\mathcal T]$ and outgoing-radiation term $\mathcal R(\mathcal T)$ in Eq.~\eqref{eq:gsebm} do not contribute. For W$\to$SB, the negative forcing therefore cools high latitudes strongly because their heat capacity $\mathcal C(x)$ is small. Heat transport resumes after the jump and redistributes this temperature imbalance, producing the loop in the projected path in Fig.~\ref{fig:gsebm}(a). For type C with $\alpha=3/2$ in the W$\to$SB transition, we further checked that the jump persists under regularization, while replacing it by a finite-duration kick increases the action by $0.44\%$ for one month and $5.1\%$ for one year. The jump accounts for most of the type C action, see Fig.~\ref{fig:gsebm}(c). Overall, the transition actions are again lower than for Gaussian noise, particularly for types B and C (cf.\ Table~\ref{tab:gsebm} and Fig.~\ref{fig:gsebm}(c)). We note that for types B and C, the effective irradiance $\mu + \eta$ becomes negative over short intervals, and the type C jump removes the equivalent of about one year of mean insolation. This is not physical as a literal solar irradiance, and we refer to Ref.~\cite{lucarini-serdukova-margazoglou:2022} for a discussion.

\paragraph*{Monte Carlo.}
Following Ref.~\cite{lucarini-serdukova-margazoglou:2022}, we apply noise increments as constant forcing over intervals $\Delta t$, a scheme converging to the Marcus SDE as $\Delta t\to0$~\cite{kurtz-pardoux-protter:1995}. The forcing intervals must resolve the time scale of the instanton control: annual intervals suffice for Gaussian and canonical type A noise, whose forcing is spread over several years. We also test a modified setup with higher jump rates $r = 0.64$ below, as well as monthly instead of yearly forcing intervals for the Monte Carlo simulations. We extract transition paths from long stationary runs and compare them with the instantons in Fig.~\ref{fig:gsebm-mc} (see Appendix~\ref{app:gsebm-sampling} for details).

For Gaussian noise, the mean transition paths agree well with the instantons, with discrepancies small relative to the path spread and shrinking as $\eps$ decreases, see Fig.~\ref{fig:gsebm-mc}(a,e). Here, at $\eps=0.49$, the canonical forcing amplitude is $\sqrt{0.04\eps}=0.14$, as in Ref.~\cite{lucarini-serdukova-margazoglou:2022} (up to possible differences in the normalization convention used for their $\alpha$-stable noise).

For types A/B at the canonical intensity $r=0.01$, sampled transition paths remain different from the instantons at the accessible noise strengths, see Fig.~\ref{fig:gsebm-mc}(i,j). For type A in Fig.~\ref{fig:gsebm-mc}(i), the discrepancies at this low intensity indicate slow convergence toward the instanton. Reaching the regime of many small jumps that our LDP assumes would require smaller $\eps$, which is out of reach for direct Monte Carlo. For type B in Fig.~\ref{fig:gsebm-mc}(j), sampled transitions are dominated by single kicks near the smallest impulse crossing the separatrix, resembling the critical-kick mechanism studied for $\alpha$-stable noise in Ref.~\cite{lucarini-serdukova-margazoglou:2022}. Integrating the type B tilted rate $r/[2\eps(1-a^2u(t)^2)]$, with $u=\sigma(\phi_z)^\top p_z$, over the forcing pulse gives only about $0.065/\eps$ jumps. The discrepancy therefore again reflects a finite-noise regime with too few jumps (rather than insufficient forcing resolution; annual controls change the type B instanton only slightly).

To access the many-small-jump regime, we hence increase the intensity to $r=0.64$ and reduce the amplitude scale to $a=0.25$, preserving the variance, and recompute the type A/B instantons. At the higher jump rate, sampled transitions involve many small jumps, and the transition paths agree more closely with the corresponding instantons as $\eps$ decreases, see Fig.~\ref{fig:gsebm-mc}(b,c,f,g). Further discussion of the critical kicks is given in Appendix~\ref{app:gsebm-sampling}.

For canonical type C ($\alpha=1/2$), a dominant monthly increment reproduces the jump mechanism of the instanton, see Fig.~\ref{fig:gsebm-mc}(d,h). Its exponential jump tail, of scale $\eps a$, causes an overshoot of the instanton landing point that dominates the path discrepancy. The sampled paths agree more closely with the instanton as $\eps$ decreases, but smaller $\eps$ yields too few transitions for a reliable comparison within feasible simulation times. We represent the sampled transitions for type C and canonical type B noise by their median path, aligned at the largest monthly increment, rather than by the conditional mean.

We do not compute prefactors for this example, since the Monte Carlo data concern transitions in long stationary runs, and a quantitative comparison would require transition rates, i.e., the infinite-time limit together with an Eyring--Kramers type formula~\cite{bouchet-reygner:2016}, which is beyond the scope of this work.

%%%%%%%%%%%%%%%%%%%%%%%%%%%%%%%%%%%%%%%%%%%%%%%%%%%%%%%%%%%%%%%%%%%%%%%%%%%

\section{Discussion and outlook}
\label{sec:conclusion}
%%%%%%%%%%%%%%%%%%%%%%%%%%%%%%%%%%%%%%%%%%%%%%%%%%%%%%%%%%%%%%%%%%%%%%%%%%%

We have shown that the OC formulation of sample path large deviations carries over from Gaussian noise to light-tailed L{\'e}vy noise in the regime of small jumps at high intensity, with the Legendre transform of the noise CGF as running cost. This makes it possible to reuse the same numerical machinery, including automatic differentiation, for instanton computations across noise types, in systems ranging from low-dimensional SDEs to a spatially extended climate model with multiplicative noise. For continuous instantons, the prefactor of the sharp large deviation estimate is obtained through the same Riccati or Fredholm determinant expressions as in the Gaussian case, with the constant noise covariance replaced by the covariance $\nabla^2 \Psi(p_z)$ of the exponentially tilted noise along the instanton. In contrast to the Gaussian case, the prefactor generically depends on the endpoint already for one-dimensional gradient systems, and replacing it by its Gaussian counterpart can misestimate tail probabilities. For continuous, non-degenerate instantons, the resulting sharp estimates agreed well with Monte Carlo simulations in the two-dimensional model and for the RTP.

A new feature compared to Gaussian noise is the possibility of discontinuous instantons, which are admissible whenever the CGF has a bounded domain, but optimal only if the subleading behavior of the cost and the drift allow for it. The optimality conditions of the TR problem~\eqref{eq:tr-problem} show that type B instantons never jump, while type C instantons can, which generalizes the one-dimensional picture of Ref.~\cite{baule-sollich:2023}. In the examples, type C instantons jumped in the two-dimensional model and in the EBM. Type B instantons remained continuous, with a short pulse of the forcing in the latter, whereas for the RTP, type C instantons contained saturated arcs. TR, with physical time frozen during jumps, provided a robust way to compute such instantons. The RTP and the EBM further showed that, for L{\'e}vy noise with low jump intensity, the small-noise regime described by the instanton is only reached once a rare event is built from many noise jumps, which can require much smaller values of $\eps$ than for Gaussian noise.

Several questions remain open. First, the prefactor theory for jumping or saturated instantons is incomplete. Our splitting heuristic worked only for integrable singularities of $\nabla^2 \Psi$ at the jump and away from the jump onset, and a uniform theory through the onset of a jump, as well as a treatment of the non-Gaussian fluctuations in the saturated regime, are left for future work. Second, all instantons here were computed on finite time horizons, and for the transitions in the EBM, we approximated the infinite-time escape from the source state by a finite control window. For transitions between metastable states, the natural next step is to combine the TR with an infinite-time formulation, i.e., gMAM for L{\'e}vy noise, in which the escape from and approach to the fixed points, as well as the prefactor, are treated by linearization around the fixed points~\cite{shao-grafke-mackay:2026}. Third, importance sampling based on the computed instantons, which we used here for compound Poisson tumbles of the RTP, would allow testing the asymptotics in further regimes inaccessible to direct sampling, such as saturated turns of the RTP with infinite-activity type C noise, or transitions in the climate model at values of $\eps$ small enough that transitions are built from many noise jumps. Finally, it would be interesting to extend our approach to stochastic PDEs driven by space-time L{\'e}vy noise, and to the crossover to heavy-tailed noise or jump rates $r \to 0$, where large single jumps dominate.

\begin{acknowledgments}
The authors thank Karthik Cheruvary, Rafael D{\'i}az Hern{\'a}ndez Rojas, and Peter Sollich for helpful discussions and for sharing their insights and unpublished results on the RTP.
The authors further thank Tobias Grafke and Georg Stadler for useful discussions and comments.
The authors acknowledge use of generative AI to assist in derivations, coding, and manuscript writing. All AI output was checked by the
authors, and they take full responsibility for the results presented in this work.
\end{acknowledgments}

\section*{Author declarations}

\subsection*{Conflict of interest}

The authors have no conflicts to disclose.

\section*{Data availability}

\vspace{-.4cm}

The data that support the findings of this study are available on request from the corresponding author. Example code for the two-dimensional example can be found under \url{https://github.com/TimoSchorlepp/sharp-extreme-event-levy}.

\appendix 

\section{It{\^o} versus Marcus}
\label{app:ito-marcus}

We recall that a multiplicative noise term $\sigma(X) \, \dd Z$ requires an interpretation of the stochastic integral. For Brownian noise, the It{\^o} and Stratonovich interpretations evaluate $\sigma$ at the beginning and at the midpoint of each time step, respectively, and differ by a noise-induced drift. For a noise jump $\Delta Z$ at time $t$, the It{\^o} interpretation evaluates $\sigma$ at the state $X_{t^-}$ before the jump, such that the state moves along the straight segment $X_t = X_{t^-} + \sigma(X_{t^-}) \Delta Z$. The Marcus interpretation, the natural generalization of the Stratonovich interpretation to jumps, instead resolves the jump by following the flow of the noise vector field for a unit of fictitious time, $X_t = y(1)$ with $y' = \sigma(y) \Delta Z$, $y(0) = X_{t^-}$. Like Stratonovich SDEs, Marcus SDEs obey the ordinary chain rule, and they arise as limits of smooth approximations of the noise, e.g., of forcing that is constant over short time intervals (Wong--Zakai approximation)~\cite{marcus:1981,kurtz-pardoux-protter:1995}. Expanding the flow gives $y(1) = X_{t^-} + \sigma \Delta Z + \frac12 [(\sigma \Delta Z) \cdot \nabla] \sigma \, \Delta Z + O(|\Delta Z|^3)$, with $\sigma$ and its derivatives evaluated at $X_{t^-}$, such that the two interpretations differ at second order in the jump size. For discretizations, this means the following: an explicit Euler step $X_{k+1} = X_k + b(X_k) \Delta t + \sigma(X_k) \Delta Z_k$ is consistent with the It{\^o} SDE even if the increment $\Delta Z_k$ contains a large jump, whereas a scheme for the Marcus SDE has to apply the increment along the flow of $\sigma$, e.g., by integrating $y' = b(y) \Delta t + \sigma(y) \Delta Z_k$ over a unit of fictitious time with an ODE solver. The difference is negligible for small jumps, but of order one for large ones.

\section{Prefactors for several terminal constraints}
\label{app:multi-constraint}

We state the generalization of the prefactor expressions of Sec.~\ref{sec:fredholm} to the joint PDF $\rho_\eps(z)$ of a vector-valued observable $f = (f_1, \dots, f_\ell) \colon \RR^n \to \RR^\ell$ at $z \in \RR^\ell$, as used for the RTP in Sec.~\ref{sec:rtp}. We have
\begin{align}
\rho_\eps(z) \overset{\eps \downarrow 0}{\sim} (2 \pi \eps)^{-\ell/2} \, C(z) \exp \left\{ -\frac{I(z)}{\eps}\right\}\,,
\label{eq:multi-sharp}
\end{align}
where now $\lambda_z \in \RR^\ell$ and the first-order condition~\eqref{eq:first-order} reads $\nabla L(\eta_z) = \sum_{j=1}^\ell \lambda_{z,j} \delta F_j / \delta \eta [\eta_z]$. Denote by $G \in \RR^{\ell \times \ell}$ the Gram matrix of the whitened constraint gradients, $G_{ij} = \langle M_z \delta F_i / \delta \eta, M_z \delta F_j / \delta \eta \rangle_{L^2}$, and by $\ppr_z$ the $L^2$-orthogonal projection onto the complement of their span. Then, in the Fredholm formulation,
\begin{align}
C(z) &= \left[ \det G \, \det \left(\Id - A_z\right) \right]^{-1/2}\,, \nonumber\\
A_z &:= \ppr_z M_z \left(\sum_{j=1}^\ell \lambda_{z,j} \frac{\delta^2 F_j}{\delta \eta^2}[\eta_z]\right) M_z \ppr_z\,,
\label{eq:multi-fredholm}
\end{align}
which reduces to Eq.~\eqref{eq:fredholm-prefactor} for $\ell = 1$, since then $\det G = \norm{M_z p_z}_{L^2}^2 / \lambda_z^2$. In the Riccati formulation, with $B := \nabla f(\phi_z(T)) \in \RR^{\ell \times n}$, $\Lambda := \sum_{j = 1}^\ell \lambda_{z,j} \nabla^2 f_j(\phi_z(T))$ and $Q$ the solution of Eq.~\eqref{eq:riccati},
\begin{align}
C(z) &= \biggl[\det\left(\Id - Q(T) \Lambda\right) \nonumber\\
&\qquad \times \det \left( B \left(\Id - Q(T) \Lambda\right)^{-1} Q(T) B^\top \right)\biggr]^{-1/2} \nonumber\\
&\quad \times \exp\left\{\frac{1}{2} \int_0^T \trace\left[\nabla^2_x \langle p_z, b(\phi_z) \rangle \, Q \right] \dd t\right\}\,.
\label{eq:multi-riccati}
\end{align}
This follows from the sharp asymptotics $(2 \pi \eps)^{-n/2} \det Q(T)^{-1/2} \exp\{\frac12 \int_0^T \trace[\nabla^2_x \langle p_z, b(\phi_z)\rangle Q] \dd t - S(x)/\eps\}$ of the PDF of $X^\eps_T$ at $x$, with $\nabla^2 S(\phi_z(T)) = Q(T)^{-1}$~\cite{schorlepp-grafke-grauer:2021,grafke-schaefer-vanden-eijnden:2023}, by a Laplace approximation of the integral over the level set $\{f = z\}$. For a linear observable $f(x) = B x$, we have $\Lambda = 0$, and Eq.~\eqref{eq:multi-riccati} reduces to $C(z) = \det(B Q(T) B^\top)^{-1/2} \exp\{\dots\}$, which is Eq.~\eqref{eq:riccati-prefactor} for $\ell = 1$.

\section{Numerical details}
\label{app:regularize}

This appendix collects some of the numerical details needed to reproduce the results reported in the main text.

\subsection{Cutoff in $\kappa$ for the perspective function}

In practice, we want to avoid numerical instabilities close to $\kappa = 0$ in Eq.~\eqref{eq:perspective}.
For the two-dimensional example problem of Sec.~\ref{sec:2d}, we approximate $\bar{L}$ by $L_\infty$ when $\kappa < 10^{-5}$ to avoid division by small numbers; for types B and C, $L_\infty(\hat{\eta}) = \sum_i |\hat{\eta}_i| / a_i$. This cutoff discards subleading small-$\kappa$ terms, so we identify jumps by finite state motion at frozen physical time and not by the cutoff alone. In the EBM, we evaluate the perspective directly for $\kappa > 0$ and use the recession cost only at $\kappa = 0$.

\subsection{Regularization of the time reparameterization}
\label{app:regularize-details}

In contrast to gMAM, we do not impose an arclength parameterization. Note that besides genuine jumps, the TR formulation may result in intervals with $\kappa = 0$ and $q' = 0$, which are harmless padding of artificial time where the state does not move. In particular, the presence of a jump can only be inferred from the physical graph $\{(\tau(s), q(s))\}$, and not from $\kappa$ alone. To avoid wasting artificial grid points for this, as well as to regularize the TR, we impose pointwise bounds $0 \leq \kappa(s) \leq \min\{c_\kappa \norm{q'(s)}, \kappa_{\max}\}$, which prevent physical time from advancing while the state is stationary and limit the physical time step per grid cell. We impose $0\leq\kappa\leq\kappa_{\max}$ as box bounds in L-BFGS-B (in the two-dimensional example) and enforce the geometric bound $\kappa\leq c_\kappa\norm{q^\prime}$ through a quadratic penalty within the augmented Lagrangian method. Optionally, a weak penalty $\frac{\beta_v}{2} \int_0^1 (v - \bar{v})^2 \, \dd s$ on the deviation of the generalized speed $v = (\norm{q'}^2 + \kappa^2)^{1/2}$ from its mean $\bar{v}$ favors a uniform use of artificial time.

These choices do not yet make the problem uniquely solvable, and $\kappa$ can still oscillate on the grid scale, or vanish on intervals without state motion. We thus introduce a smoothness penalty $\frac{\beta_\kappa}{2} \int_0^1 |\kappa'(s)|^2 \, \dd s$ which helps suppress grid-scale oscillations, but does not by itself guarantee removal of padding intervals or preservation of a genuine jump. Since this penalty is not invariant under reparameterizations, it does, however, bias the physical graph if it is too strong, by smearing out part of a jump over a short physical time interval with small $\kappa > 0$. We therefore first redistribute the grid of a converged unregularized solution uniformly in the augmented arclength $\int_0^s (\tau'^2 + c_{\rm s}^2 \norm{q'}^2)^{1/2} \dd s'$, with a constant weight $c_{\rm s} > 0$, which removes the padding without changing the physical graph, and then re-optimize with a weak smoothness penalty, which favors a smooth representative while limiting the perturbation of the physical graph. The bound $\kappa \leq c_\kappa \norm{q'}$ restricts how long the instanton can rest near a fixed point, and could in principle bias the minimizer; we therefore cross-check jumping instantons independently, see Appendix~\ref{app:2d-num} below. We always report the physical action without any regularization terms.

\subsection{Details for Sec.~\ref{sec:2d}}
\label{app:2d-num}

Gaussian, type A and type B rate functions shown in the main text in Fig.~\ref{fig:2d-rates} are obtained by fixed-time control optimization of Eq.~\eqref{eq:control-problem} discretized via $1000$ explicit-Euler time steps (and we used shooting, i.e., direct solution of Eq.~\eqref{eq:hamilton} as a reference to cross-check). The type C rate functions in Fig.~\ref{fig:2d-rates} and jump-prefactor curves in Fig.~\ref{fig:2d-prob} use an initial TR run using Eq.~\eqref{eq:tr-problem} plus a subsequent shooting refinement, as described below.

Our TR computations for Eq.~\eqref{eq:tr-problem}, which we tested for types B and C, use $500$ equidistant grid cells in artificial time, with $c_\kappa = 5$, $\kappa_{\max} = 5T$ for type C, $c_\kappa = 2.5$, $\kappa_{\max} = 2T$ for type B. The type B and type C ($\alpha=1/2$) paths at $z=2,3,4$ in Fig.~\ref{fig:2d-instantons}(b) use the generalized-speed penalty $\beta_v=10^{-4}$ and no smoothness penalty ($\beta_\kappa=0$). For the type C TR solution at $z=5$ in Fig.~\ref{fig:2d-jump}, we instead set $\beta_v=0$ and, after redistributing the grid with $c_{\rm s}=0.3$, use a smoothness penalty $\beta_\kappa=10^{-4}$. At $\beta_\kappa = 10^{-4}$, some displacement near the jump already occurs at small positive $\kappa$, outside the frozen-time window. Increasing the penalty to $\beta_\kappa = 3 \cdot 10^{-3}$ produces stronger smearing and a visible shift of the horizontal part of the instanton.  In the type B computations, cells where the cutoff in $\kappa$ is reached carry only negligible state motion.

TR is our main OC formulation, which we used for all initial scans. We then used Hamiltonian shooting, with the TR result as an initial guess, to refine and independently check smooth segments and jump matching. All prefactors are computed in physical time.

\subsection{Details for Sec.~\ref{sec:rtp}}
\label{app:rtp-numerics}
\paragraph*{Optimization and prefactor evaluation.}
We minimize Eq.~\eqref{eq:rtp-reduced} over discretized angular histories with a Newton trust-region method and continuation in the endpoint. For type B tumbles, we then solve the instanton equations in continuous time with an adaptive collocation method, starting from the discrete minimizers, and evaluate $C_{\theta_0^*}$ with the Riccati formulation of Sec.~\ref{sec:fredholm} and $\partial^2_{\theta_0} I$ from instantons with fixed initial angle. This avoids the slow, first-order convergence in the time step of the discretized determinant where the instanton turns quickly and the angular momentum comes close to the boundary of $\dom \Psi_\theta$.

\paragraph*{Sampling estimator.}
In all simulations, we sample only the angular noise, evaluate the Gaussian density of $(X_T, Y_T)$ conditional on the angular history analytically, and integrate over the uniformly distributed initial angle numerically. For type B tumbles, direct sampling at small $\eps$ would almost never produce the turns of the instanton discussed in Sec.~\ref{sec:rtp}. We therefore use importance sampling, and sample the angular noise from the tilted measure~\eqref{eq:tilt}, with the tilt $p(t) = p_\theta(t)$ given by the angular momentum of the instanton for the respective endpoint. For type B noise, the tilted L{\'e}vy measure $e^{p_\theta y / \eps} \nu_\eps(\dd y)$ is again that of a compound Poisson process, since the tilt only changes the decay rates of the exponential jump size density: positive and negative tumbles arrive at rates $r / (4 \eps (1 \mp a p_\theta(t)))$, with exponentially distributed sizes of mean $\eps a / (1 \mp a p_\theta(t))$. Since the tumbles are symmetric and hence need no compensation, a sampled angular history with tumbles $\Delta \Theta_k$ at times $t_k$ carries the explicit weight
\begin{align*}
\textstyle \frac{\dd \PP}{\dd \mathbb{Q}} = \exp\biggl\{-\frac{1}{\eps} \sum_k p_\theta(t_k) \Delta \Theta_k  + \frac{1}{\eps} \int_0^T \Psi_\theta(p_\theta(t)) \, \dd t\biggr\}\,,
\end{align*}
and $\rho_\eps(x_{\rm f}, y_{\rm f})$ is estimated by the sample mean of this weight times the conditional Gaussian density of $(X_T, Y_T)$, integrated over $\Theta_0$. This estimator is unbiased for any tilt. With the instanton tilt, the sampled histories concentrate around the instanton, such that, in contrast to direct sampling, the required number of samples does not grow exponentially as $\eps \downarrow 0$. We add samples until the relative standard error is below $0.5\%$.

\paragraph*{Additional convergence checks.}
At the turning endpoints, both on the $y_{\rm f} = 4$  slice and on the slices $y_{\rm f} = 6$ and $x_{\rm f} = 6$ with the most strongly turning endpoints (not shown), the ratio of simulated and sharp PDF approaches one as $\eps$ decreases, with the observed decrease in deviation for $\eps \leq 10^{-3}$ consistent with a correction of order $\eps$. Many tumbles occur only for $\eps \lesssim 10^{-3}$, and the endpoint-conditioned mean angular histories then follow the late turn of the instanton closely.

\subsection{Details for Sec.~\ref{sec:gsebm}}
\label{app:gsebm-sampling}

\paragraph*{Control-grid resolution.}
The TR control window is discretized into $N_{\rm c}=40$ cells in artificial time. For W$\to$SB with type B noise, refinement to $N_{\rm c}=80$ and $160$ gives actions 0.54105 and 0.54100, compared with 0.5413 at $N_{\rm c}=40$. For type C with $\alpha=3/2$, refinement to $N_{\rm c}=80$ gives 0.60180, compared with 0.60185 at $N_{\rm c}=40$. These checks indicate weak grid dependence of the actions in these cases.

\paragraph*{Monte Carlo sampling.} We follow the Monte Carlo approach of Lucarini et al.~\cite{lucarini-serdukova-margazoglou:2022}, in which the noise increments are applied as forcing that is constant over intervals of length $\Delta t$, i.e., the state follows the deterministic flow with this forcing during each interval. This Wong--Zakai-type scheme converges to the Marcus SDE as $\Delta t \to 0$~\cite{kurtz-pardoux-protter:1995}, see Appendix~\ref{app:ito-marcus} (although formulated in Ref.~\cite{lucarini-serdukova-margazoglou:2022} in the It{\^o} sense, we stress that piecewise-constant forcing integrated by an ODE
solver indeed converges to the Marcus SDE), and, in particular, a large increment moves the state along the flow of the noise vector field rather than along a straight It{\^o} update, consistent with the instantons. As $\eps \downarrow 0$, such simulations follow the instanton with controls that are piecewise constant on these intervals rather than the continuous-time instanton, so $\Delta t$ has to resolve the time scale of the instanton forcing. By default, we use $\Delta t = 1$ year as in Ref.~\cite{lucarini-serdukova-margazoglou:2022}; this suffices for Gaussian and type A noise, whose instanton forcing is spread over several years. For the high-rate type A and type B laws, we use $\Delta t = 1$ month, together with the increased jump rate as explained in the main text to access transitions built from many small jumps.

Monthly forcing alone does not bring canonical type B sampling into the ``many-small-jumps'' regime. For type C, we also use monthly forcing, replacing the instanton jump by a one-month kick. We extract all transitions from long stationary runs and summarize them by a mean transition path: the mean of $\Delta \mathcal{T}$ conditional on $\langle \mathcal{T} \rangle$ for continuous instantons, and, for transitions that are triggered by a single large monthly increment, the median over transitions aligned at this increment, from one year before to two years after it. Figure~\ref{fig:gsebm-mc} compares these paths to the instantons.

\paragraph*{Canonical type B checks.}
For canonical type B noise, simulations with annual forcing over a range of $\eps$ (not shown) give transition counts close to the probability that one annual increment exceeds the critical annual kick. At $\eps=0.08$, the largest monthly increment in every observed transition contains a single noise jump, usually comparable to the critical one-month kick. The corresponding instanton contains no jump. These checks support the finite-noise interpretation of the discrepancy in Sec.~\ref{sec:gsebm}. The type B instanton changes only slightly when controls are constrained to be constant over each year, showing that monthly forcing alone cannot remove the discrepancy.

\section{Derivations of the Fredholm prefactor}
\label{app:fredholm}

\subsection{Exponential tilt}
\label{sec:fredholm-tilt}

Inserting $\dot{Z}^\eps = \eta_z + \sqrt{\eps} M_z \xi$ into Eq.~\eqref{eq:tilt} and using the Legendre identity $\langle p_z, \eta_z \rangle - \Psi(p_z) = L(\eta_z)$, the likelihood ratio becomes
\begin{align*}
\frac{\dd \PP}{\dd \mathbb{Q}} = e^{-I(z)/\eps} \exp\left\{-\eps^{-1/2} \langle M_z p_z, \xi \rangle_{L^2}\right\}\,,
\end{align*}
which is linear in $\xi$ in the exponent.
Now, $\rho_\eps(z) = \EE_{\mathbb{Q}}[\delta(F - z) \, \dd \PP / \dd \mathbb{Q}]$. Denote by $u_z := M_z \, \delta F / \delta \eta [\eta_z] = M_z p_z / \lambda_z$ the whitened gradient, where we used $p_z = \lambda_z \delta F / \delta \eta$ from Eq.~\eqref{eq:first-order}, and decompose $\xi = \xi_\parallel \hat{u}_z + \xi_\perp$ with $\hat{u}_z = u_z / \norm{u_z}_{L^2}$ and $\langle \hat{u}_z, \xi_\perp \rangle_{L^2} = 0$. Expanding
\begin{align*}
F[\eta_z + \sqrt{\eps} M_z \xi] &= z + \sqrt{\eps} \norm{u_z}_{L^2} \xi_\parallel \nonumber\\
&\quad + \frac{\eps}{2} \left\langle \xi, M_z \frac{\delta^2 F}{\delta \eta^2} M_z \xi \right\rangle_{L^2} + O(\eps^{3/2})\,,
\end{align*}
the delta function becomes, to leading order,
\begin{align*}
&\delta(F - z) = \frac{\delta(\xi_\parallel - \xi_\parallel^*)}{\sqrt{\eps} \norm{u_z}_{L^2}}\,, \nonumber\\
&\xi_\parallel^* = -\frac{\sqrt{\eps}}{2 \norm{u_z}_{L^2}} \left\langle \xi_\perp, M_z \frac{\delta^2 F}{\delta \eta^2} M_z \xi_\perp \right\rangle_{L^2} + O(\eps)\,.
\end{align*}
On this set, the exponent of the likelihood ratio is $-\eps^{-1/2} \lambda_z \norm{u_z}_{L^2} \xi_\parallel^* = \frac{\lambda_z}{2} \langle \xi_\perp, M_z \frac{\delta^2 F}{\delta \eta^2} M_z \xi_\perp \rangle_{L^2} + O(\eps^{1/2})$, i.e., of order one, and the standard Gaussian density of $\xi_\parallel$ contributes $(2 \pi)^{-1/2}$ at $\xi_\parallel^* \to 0$. Hence,
\begin{align*}
\rho_\eps(z) &\overset{\eps \downarrow 0}{\sim} \frac{e^{-I(z)/\eps}}{\sqrt{2 \pi \eps} \, \norm{u_z}_{L^2}} \nonumber\\
&\quad \times \EE\left[ \exp \left\{ \tfrac{\lambda_z}{2} \textstyle \left\langle \xi_\perp, M_z \frac{\delta^2 F}{\delta \eta^2} M_z \xi_\perp \right\rangle_{L^2} \right\} \right]\,,
\end{align*}
where $\norm{u_z}_{L^2} = \norm{M_z p_z}_{L^2} / |\lambda_z|$, and, assuming $\operatorname{Id} - A_z > 0$, the expectation over the Gaussian white noise $\xi_\perp$ on the complement of $\hat{u}_z$ gives the Fredholm determinant $\det(\Id - A_z)^{-1/2}$, with $A_z$ defined in Eq.~\eqref{eq:fredholm-prefactor}, exactly as in the Gaussian case~\cite{schorlepp-tong-grafke-etal:2023}, but with $\delta F / \delta \eta$ and $\delta^2 F / \delta \eta^2$ replaced by their whitened versions.

\subsection{Path integral derivation}
\label{sec:fredholm-PI}

We sketch a formal path integral derivation of Eq.~\eqref{eq:sharp-asymptotics} with prefactor~\eqref{eq:fredholm-prefactor}, which closely follows the Gaussian case~\cite{schorlepp-tong-grafke-etal:2023}. Writing the path density of $X^\eps$ via a functional Fourier transform and using Eq.~\eqref{eq:scaled-cgf}, the PDF of $f(X^\eps_T)$ becomes
\begin{align}
\rho_\eps(z) = \int_{\phi(0) = x_0} \mathcal{D}\phi \int \mathcal{D}p \; \delta(f(\phi(T)) - z) \; J[\phi] \nonumber\\
\times \exp \left\{ -\tfrac{1}{\eps} \textstyle \int_0^T \left[ \langle i p, \dot{\phi} - b(\phi) \rangle - \Psi(ip) \right] \dd t\right\}\,,
\label{eq:path-integral}
\end{align}
where $J[\phi]$ is the $\eps$-independent Jacobian of the map $\phi \mapsto \dot{\phi} - b(\phi)$. In the Gaussian case, the $p$-integral can be carried out exactly. Here, we instead evaluate it by a saddle point approximation pointwise in time, which recovers the action $\int_0^T L(\dot{\phi} - b(\phi)) \, \dd t$ together with prefactor $\propto \prod_t \det[\nabla^2 \Psi(p^*(t))]^{-1/2}$, where $p^* = \nabla L(\dot{\phi} - b(\phi))$. Changing variables from $\phi$ to $\eta = \dot{\phi} - b(\phi)$, which cancels~$J$, we obtain
\begin{align}
\rho_\eps(z) &\overset{\eps \downarrow 0}{\sim} \int \mathcal{D}\eta \; \delta(F[\eta] - z) \prod_t \det\left[\nabla^2 \Psi(\nabla L(\eta(t)))\right]^{-1/2} \nonumber\\
&\quad \times \exp\left\{-\tfrac{1}{\eps} \textstyle \int_0^T L(\eta) \, \dd t\right\}\,.
\label{eq:eta-integral}
\end{align}
We now expand around the instanton noise, $\eta = \eta_z + \sqrt{\eps} \, \delta \eta$, and represent the delta function by an integral over a Lagrange multiplier $\lambda$. Due to the Legendre identity $\nabla^2 L(\eta) = [\nabla^2 \Psi(\nabla L(\eta))]^{-1}$, the second variation of the action, $\langle \delta \eta, \nabla^2 L(\eta_z) \delta \eta \rangle$, and the divergent product in Eq.~\eqref{eq:eta-integral} are simultaneously normalized by the change of variables $\delta \eta = M_z \, \delta \tilde{\eta}$,
which whitens the local fluctuations of the noise around the instanton. The remaining integral over $\delta \tilde{\eta}$ is then identical to the one in the Gaussian case~\cite{schorlepp-tong-grafke-etal:2023}, with the first and second variations $\delta F / \delta \eta$ and $\delta^2 F / \delta \eta^2$ replaced by their whitened versions $M_z \, \delta F / \delta \eta$ and $M_z \, \delta^2 F / \delta \eta^2 \, M_z$, leading to Eq.~\eqref{eq:fredholm-prefactor} in the main text.

\section{Prefactor of jumping instantons}
\label{app:jump-prefactor}

We give details on the Laplace evaluation of the splitting Eq.~\eqref{eq:jump-splitting} that leads to Eq.~\eqref{eq:jump-prefactor}. We assume an isolated matching point $(t_{\rm j}^*,q_-^*,q_+^*)$ in the interior, with $0<t_{\rm j}^*<T$ and $q_+^*>q_-^*$, at which the total action $S = S_1(q_-, t_{\rm j}) + (q_+ - q_-) / a + S_2(q_+, t_{\rm j})$ has a positive definite Hessian. Here, $S_1(q, t)$ is the action to reach $q$ at time $t$, and $S_2(q, t)$ is the remaining action from $q$ at time $t$ to the target, which satisfy the Hamilton--Jacobi relations $\partial_q S_1 = p$, $\partial_t S_1 = -H$ and $\partial_q S_2 = -p$, $\partial_t S_2 = H$. Hence, $\partial_{q_-} S = p_- - 1/a$, $\partial_{q_+} S = 1/a - p_+$ and $\partial_{t_{\rm j}} S = H(q_+, p_+) - H(q_-, p_-)$, where $p_\pm$ denote the momenta on either side of the jump, and the first-order conditions are the Weierstrass--Erdmann conditions $p_- = p_+ = 1/a$ and $H(q_-, p_-) = H(q_+, p_+)$ of Sec.~\ref{sec:jumps}.

The Gaussian fluctuations around the matching point are governed by the Hessian $\nabla^2 S$. Differentiating the action once more along the instanton, e.g., $\partial_t \partial_q S_1 = \dot{p}_- - S_1'' \dot{q}_-$ from $\dot{p}_- = \frac{\dd}{\dd t} \partial_q S_1(\phi_z(t), t)$, all second derivatives of $S$ can be expressed through the curvatures $S_1'' := \partial^2_{q_-} S_1$, $S_2'' := \partial^2_{q_+} S_2$ of the two segments and the instanton velocities $\dot{q}_\pm, \dot{p}_\pm$ on either side of the jump. In the variables $(q_-, q_+, t_{\rm j})$,
\begin{gather}
\nabla^2 S =
\begin{pmatrix}
S_1'' & 0 & \dot{p}_- - S_1'' \dot{q}_-\\
0 & S_2'' & -\dot{p}_+ - S_2'' \dot{q}_+\\
\dot{p}_- - S_1'' \dot{q}_- & -\dot{p}_+ - S_2'' \dot{q}_+ & \begin{aligned} &S_1'' \dot{q}_-^2 - \dot{p}_- \dot{q}_-\\ &+ S_2'' \dot{q}_+^2 + \dot{p}_+ \dot{q}_+ \end{aligned}
\end{pmatrix}\,, \nonumber\\
\begin{aligned}
\det \nabla^2 S &= S_1'' S_2'' \left(\dot{p}_- \dot{q}_- - \dot{p}_+ \dot{q}_+ \right)\\
&\quad - S_2'' \dot{p}_-^2 - S_1'' \dot{p}_+^2\,.
\end{aligned}
\label{eq:jump-hessian}
\end{gather}
The curvature $S_1'' = Q_1(t_{\rm j})^{-1}$ of the first segment is obtained from the forward Riccati equation~\eqref{eq:riccati} on $[0, t_{\rm j}]$. The curvature $W(t) := \nabla^2_x S_2(\phi_z(t), t)$ of the remaining action, with $\delta p = -W \delta x$ for neighboring solutions, satisfies the backward Riccati equation
\begin{align}
\dot{W} &= \nabla_x^2 \langle p_z, b(\phi_z) \rangle - \nabla b(\phi_z)^\top W \nonumber\\
&\quad - W \nabla b(\phi_z) + W \nabla^2 \Psi(p_z) W
\label{eq:backward-riccati}
\end{align}
on $[t_{\rm j}, T]$, and $S_2'' = W(t_{\rm j})$. Since its terminal condition at $T$ is singular in the direction of $\nabla f$, we compute $S_2''$ from the fundamental matrix of the linearized Hamiltonian flow~\cite{schorlepp-grafke-grauer:2023} along the second segment, by solving for the initial momentum perturbation $\delta p(t_{\rm j}) = -W(t_{\rm j}) \delta x(t_{\rm j})$ that is compatible with the linearized terminal conditions $\langle \nabla f, \delta x(T) \rangle = 0$, $\delta p(T) \parallel \nabla f$ (for linear $f$). $S_1''$ is obtained analogously, which is equivalent to Eq.~\eqref{eq:riccati}. Finally, inserting $\rho^{(1)}_\eps \sim (2 \pi \eps)^{-1/2} C_1 e^{-S_1/\eps}$, $\rho^{(2)}_\eps \sim (2 \pi \eps)^{-1/2} C_2 e^{-S_2/\eps}$ and $\nu_\eps$ into Eq.~\eqref{eq:jump-splitting}, the Laplace approximation of the three-dimensional integral contributes $(2 \pi \eps)^{3/2} (\det \nabla^2 S)^{-1/2}$, which yields Eq.~\eqref{eq:jump-prefactor}.

In $n$ dimensions, for a jump of size $\Delta > 0$ in an additive noise component $i$, one integrates in Eq.~\eqref{eq:jump-splitting} over $t_{\rm j}$, the state $x_-\in\RR^n$ before the jump and $\Delta$, with $\rho^{(1)}_\eps \sim (2 \pi \eps)^{-n/2} C_1 e^{-S_1/\eps}$ the $n$-dimensional transition density and $\rho^{(2)}_\eps \sim (2 \pi \eps)^{-1/2} C_2 e^{-S_2/\eps}$ the PDF of $f(X^\eps_T)$ after the jump. The powers of $2 \pi \eps$ combine as before, and the same formal local calculation gives Eq.~\eqref{eq:jump-prefactor} under the same isolation and non-degeneracy assumptions, with $\nabla^2 S$ the $(n+2) \times (n+2)$ Hessian of $S_1(x_-, t_{\rm j}) + \Delta / a_i + S_2(x_- + \Delta e_i, t_{\rm j})$, which is again assembled from the Riccati solutions of both segments and the instanton velocities at the jump.

For $\alpha \in (1,2)$, the singularity $\nabla^2 \Psi(p_z(t)) \asymp |t - t_{\rm j}|^{\alpha - 2}$ at the jump is integrable. Numerically (cf.\ Sec.~\ref{sec:2d}) we integrate both segments in the variable $|t - t_{\rm j}|^{1/2}$, which removes this singularity for $\alpha = 3/2$, evaluate the Hessian at matching points at a small distance from the jump, and extrapolate this distance to zero.

\bibliography{bib}

\end{document}